\documentclass[notitlepage,letter,11pt]{article}
\usepackage{graphicx}
\usepackage[ruled,vlined]{algorithm2e}
\usepackage{latexsym}
\usepackage{amsmath}
\usepackage{amsfonts}

\usepackage{verbatim}
\usepackage{citesort}
\usepackage{simplemargins}
\setallmargins{1in}

\newcommand{\fcn}[3]{\ensuremath{#1 : #2 \rightarrow #3}}

\title{Spring Embedders and Force Directed Graph\\ Drawing Algorithms}

\author{Stephen G. Kobourov\\University of Arizona}

\begin{document}





\maketitle

\section*{Abstract}
Force-directed algorithms are among the most flexible methods for calculating layouts of simple undirected graphs. Also known as spring embedders, such algorithms calculate
the layout of a graph using only information contained within the
structure of the graph itself, rather than relying on domain-specific
knowledge.  Graphs drawn with these algorithms tend to be
aesthetically pleasing, exhibit symmetries, and tend to produce
crossing-free layouts for planar graphs. In this survey we consider several classical algorithms, starting from Tutte's 1963 barycentric method, and including recent scalable multiscale methods for large and dynamic graphs. 

\bigskip
\noindent{\bf Keywords}: Graph drawing, network visualization, spring embedders

\section{Introduction}
\label{fd:sec:intro}

Going back to 1963, the graph drawing algorithm of
Tutte~\cite{t-hdg-63} is one of the first force-directed graph drawing
methods based on barycentric representations.  More traditionally, the
spring layout method of Eades~\cite{Eades+1984a} and the algorithm of
Fruchterman and Reingold~\cite{fr-gdfdp-91} both rely on spring
forces, similar to those in Hooke's law. In these methods, there are
repulsive forces between all nodes, but also attractive forces between
nodes which are adjacent.  

\begin{figure}[t]
\begin{center}
\includegraphics[width=.32\textwidth]{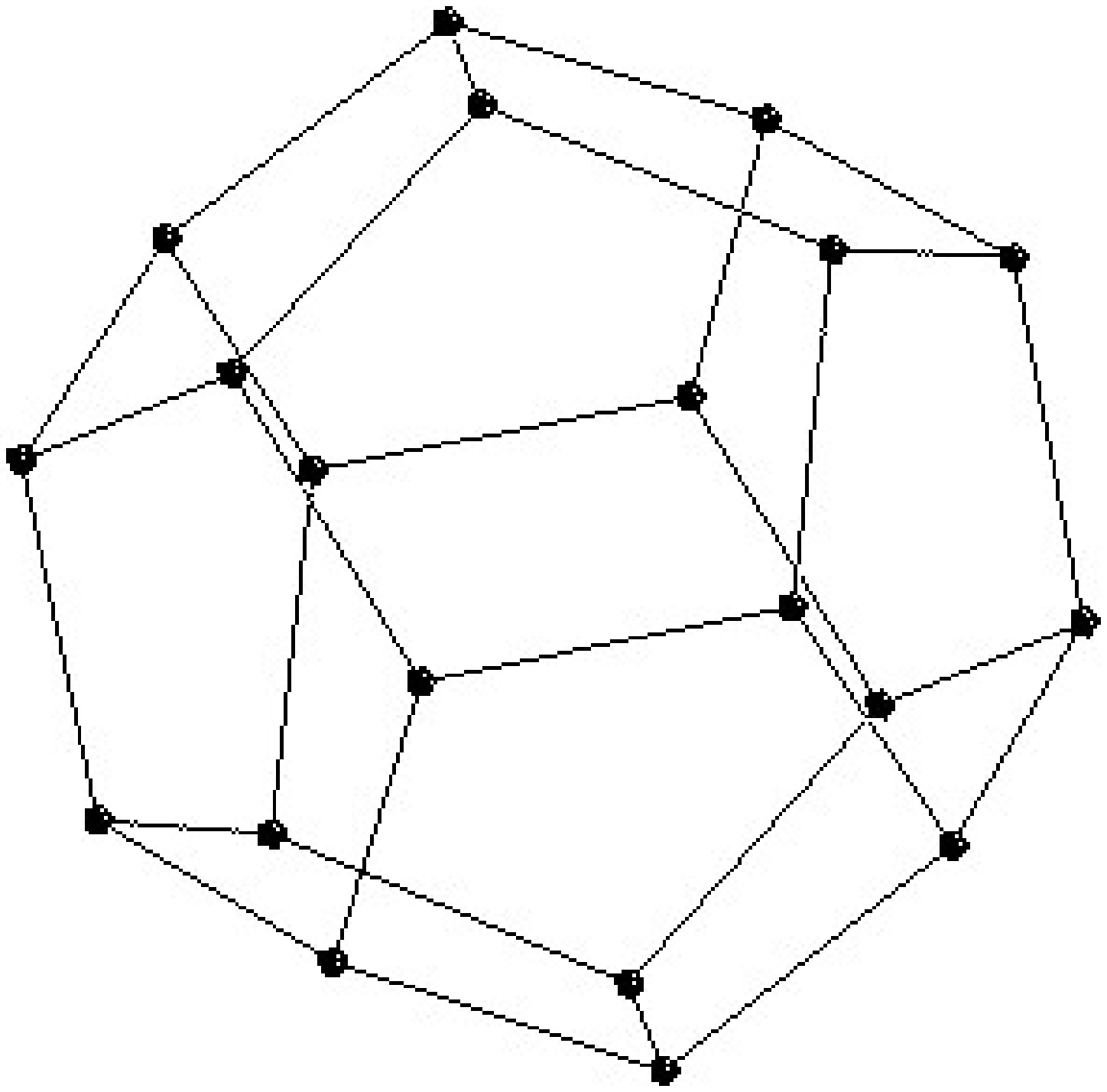}
\includegraphics[width=.32\textwidth]{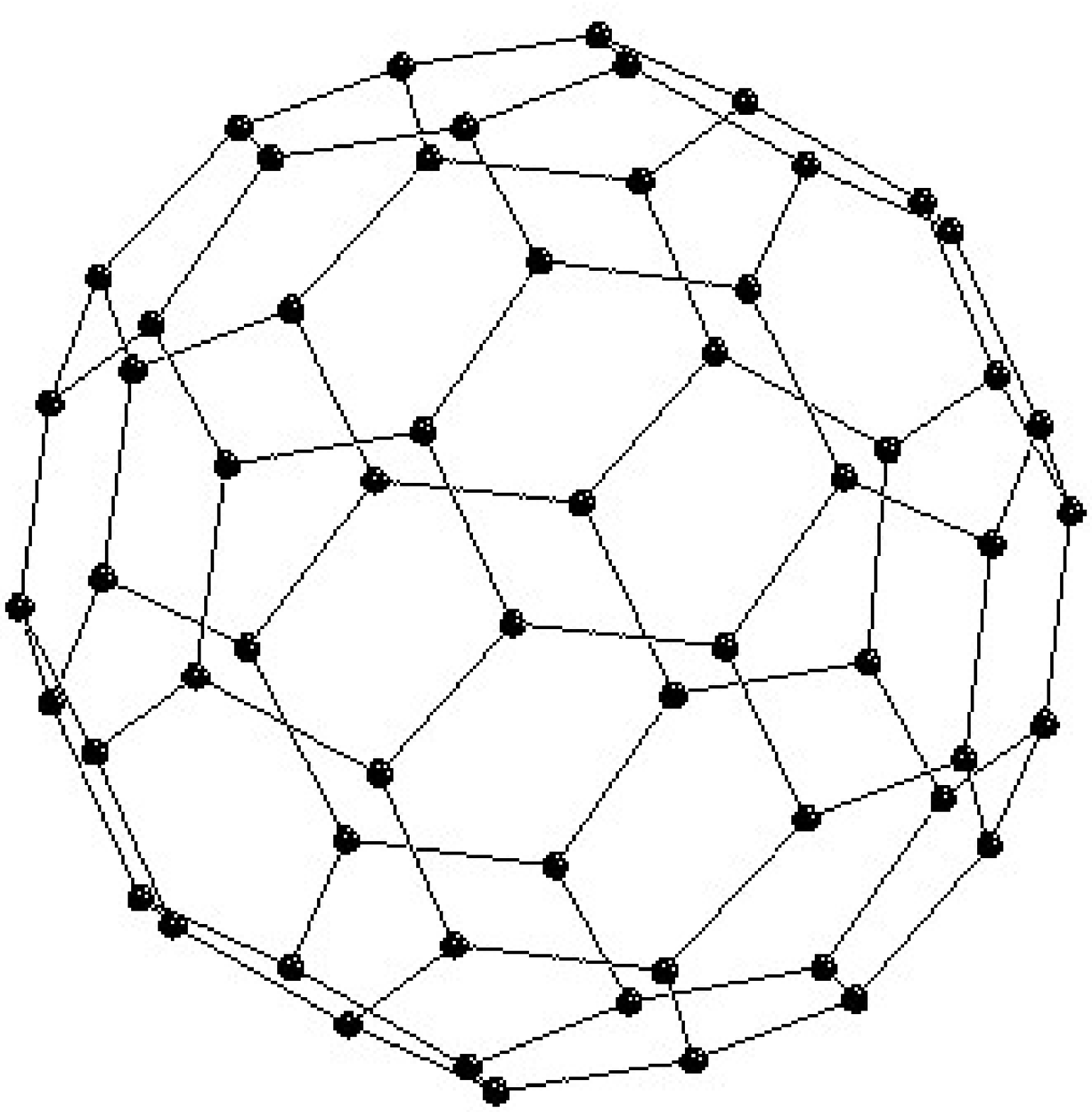}
\includegraphics[width=.32\textwidth]{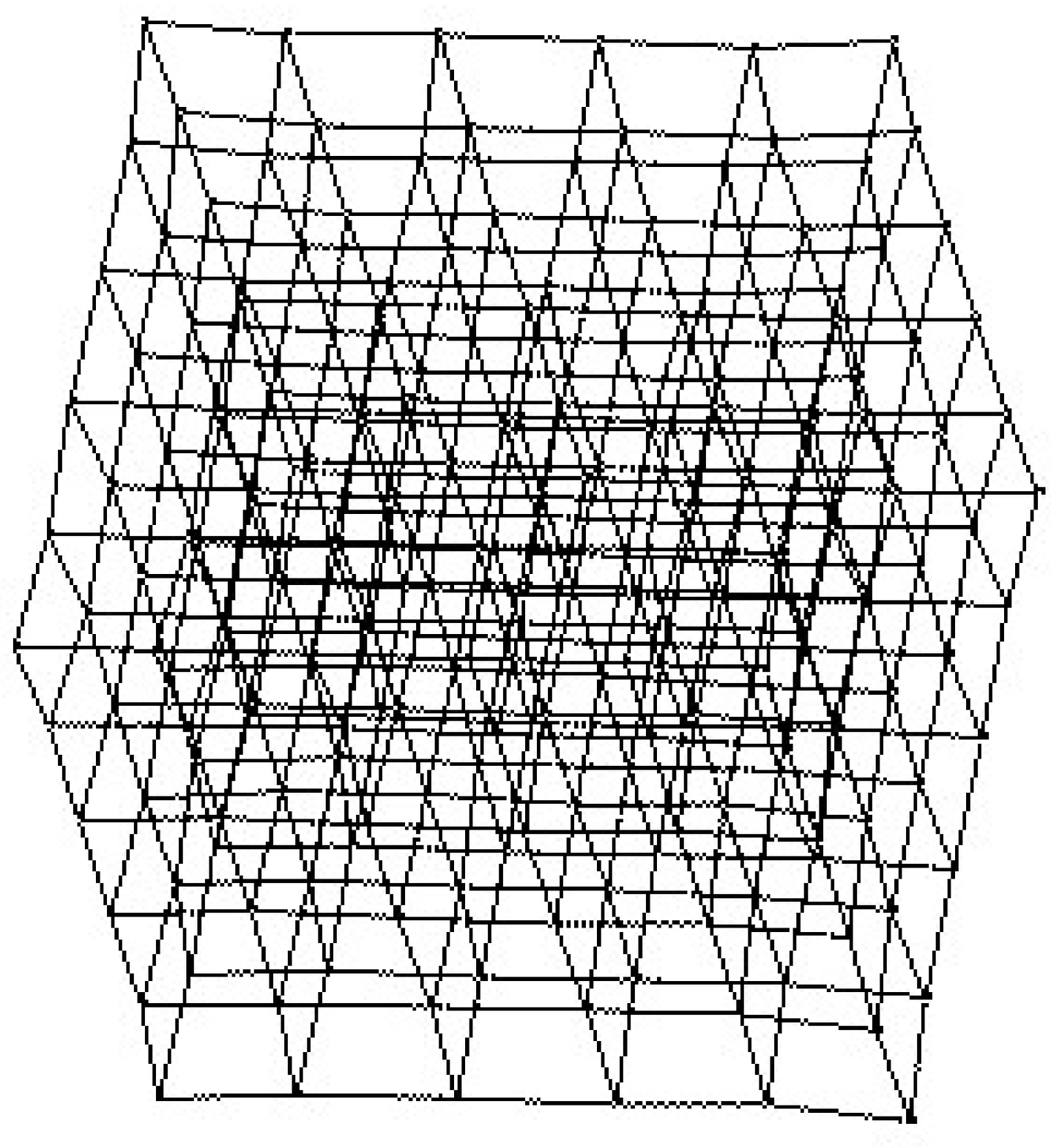}\\
\includegraphics[width=.32\textwidth]{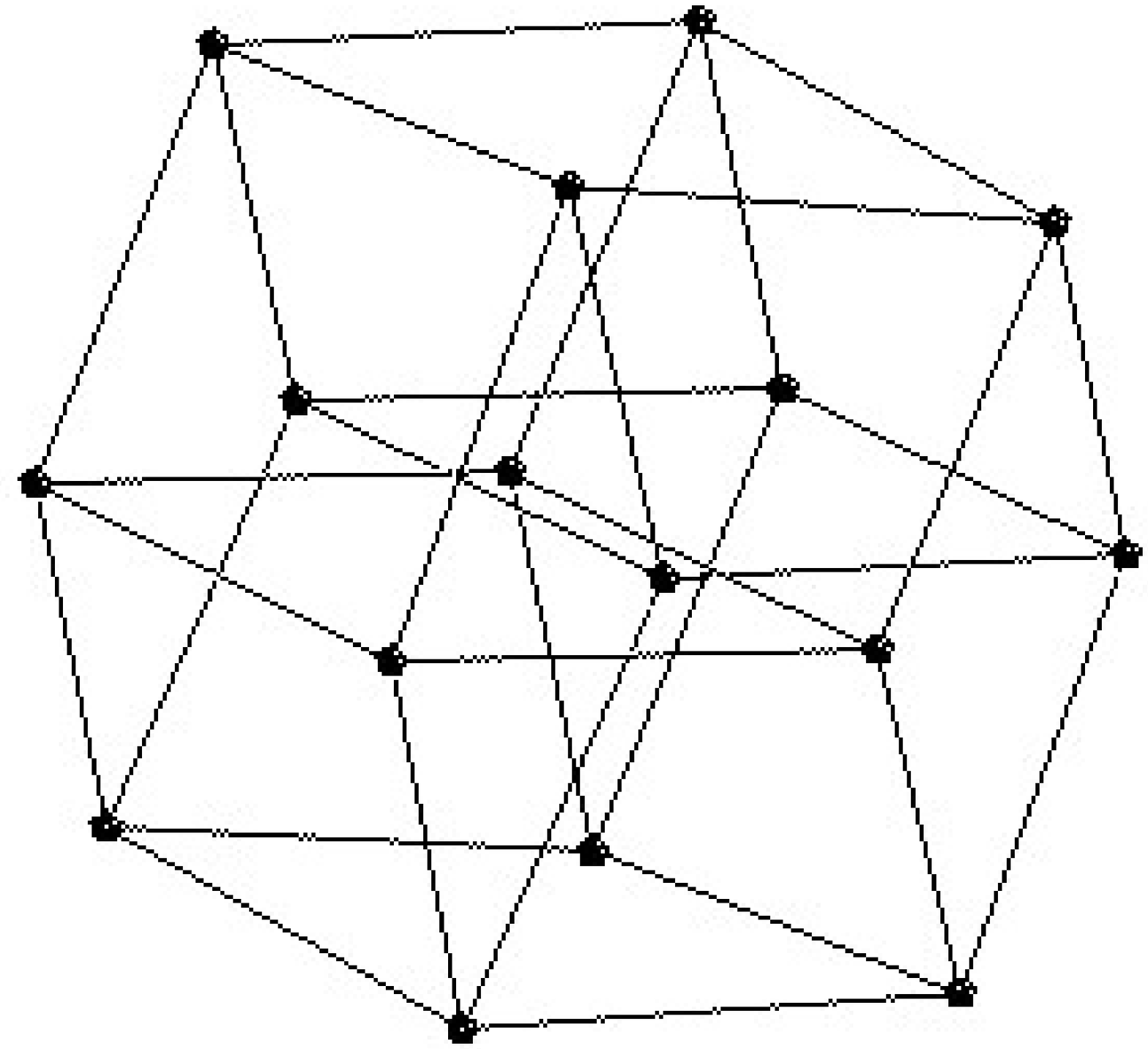}
\includegraphics[width=.32\textwidth]{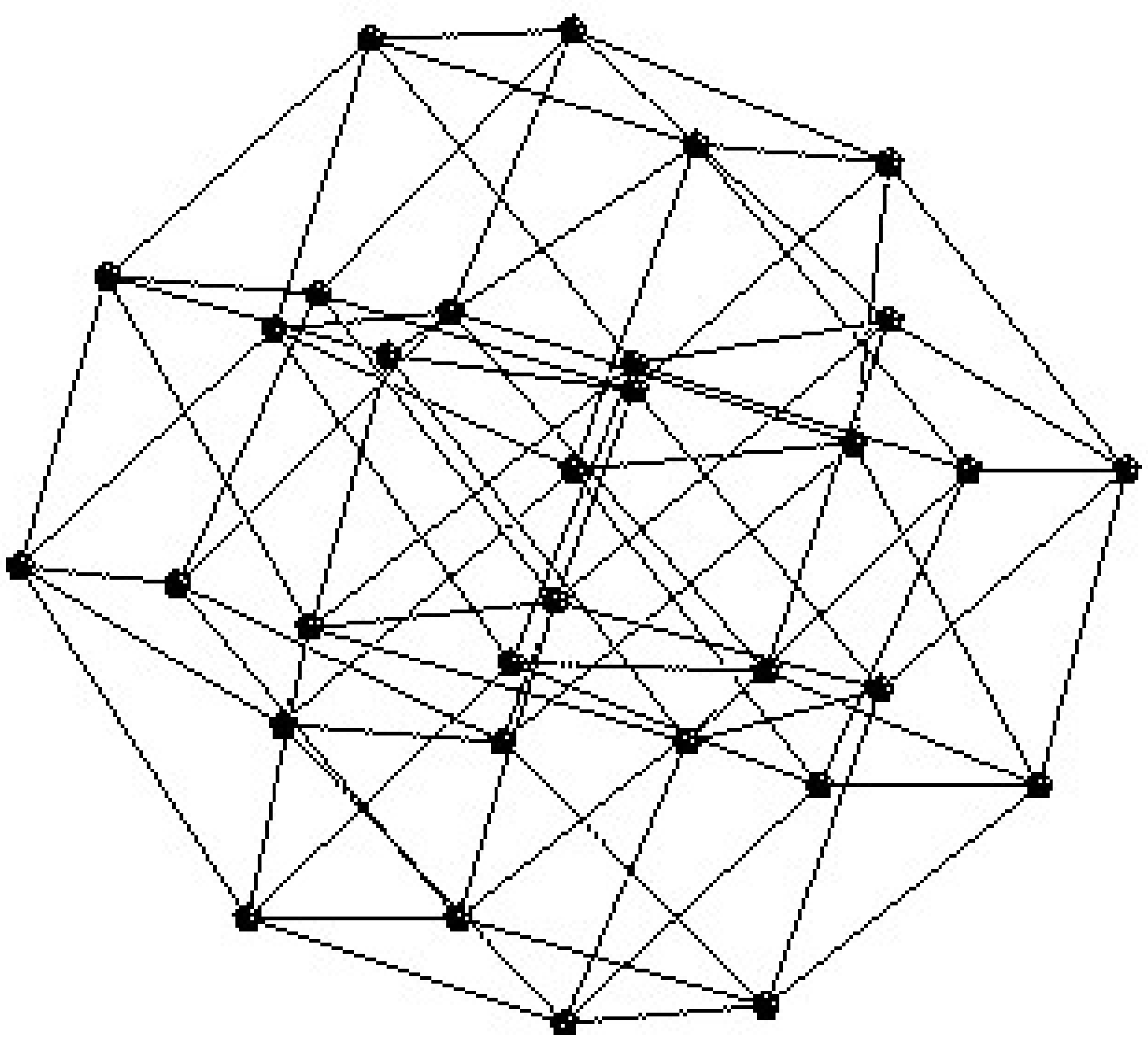}
\includegraphics[width=.32\textwidth]{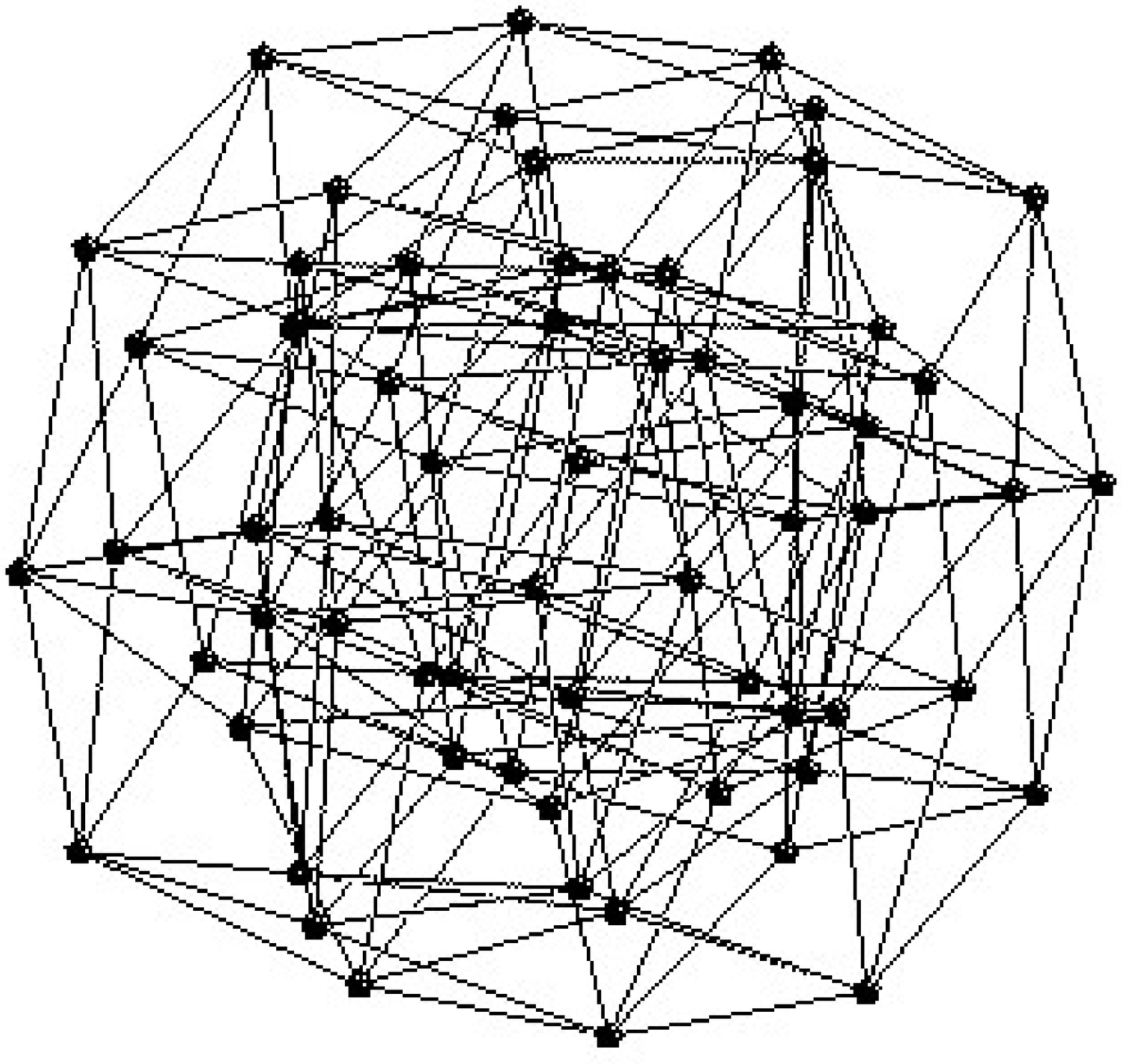}
\end{center}
\caption{\small\sf Examples of drawings obtained with a force-directed algorithm. First row: small graphs:
  dodecahedron (20 vertices), C60 bucky ball (60 vertices), 3D cube
  mesh (216 vertices). Second row: Cubes in 4, 5, 6 dimensions~\cite{gk-grip-00}.
}
\label{fd:fig:fd}
\end{figure}

Alternatively, forces between the nodes can be computed based on their
graph theoretic distances, determined by the lengths of shortest paths
between them. The algorithm of Kamada and Kawai~\cite{kk-adgug-89}
uses spring forces proportional to the graph theoretic distances. In
general, force-directed methods define an objective function which
maps each graph layout into a number in $\cal{R}^{+}$ representing the
energy of the layout.  This function is defined in such a way that low
energies correspond to layouts in which adjacent nodes are near some
pre-specified distance from each other, and in which non-adjacent
nodes are well-spaced.  A layout for a graph is then calculated by
finding a (often local) minimum of this objective function; see
Fig~\ref{fd:fig:fd}.

The utility of the basic force-directed approach is limited to small
graphs and results are poor for graphs with more than a few hundred
vertices. There are multiple reasons why traditional force-directed
algorithms do not perform well for large graphs. One of the main
obstacles to the scalability of these approaches is the fact that the
physical model typically has many local minima. Even with the help of
sophisticated mechanisms for avoiding local minima the basic
force-directed algorithms are not able to consistently produce good
layouts for large graphs. Barycentric methods also do not perform well
for large graphs mainly due to resolution problems: for large graphs
the minimum vertex separation tends to be very small, leading to
unreadable drawings.

The late 1990's saw the emergence of several techniques extending the
functionality of force-directed methods to graphs with tens of
thousands and even hundreds of thousands of vertices. One common
thread in these approaches is the multi-level layout technique, where
the graph is represented by a series of progressively simpler
structures and laid out in reverse order: from the simplest to the
most complex. These structures can be coarser graphs (as in the
approach of Hadany and Harel~\cite{hh-msadg-99}, Harel and
Koren~\cite{hk-fmsmd-j-02}, and Walshaw~\cite{w-mafdgd-j-03}, or
vertex filtrations as in the approach of Gajer {\em et al.}~\cite{ggk-afmda-00j}.

The classical force-directed algorithms are restricted to calculating
a graph layout in Euclidean geometry, typically ${\cal{R}}^2$,
${\cal{R}}^3$, and, more recently, ${\cal{R}}^n$ for larger values of
$n$. There are, however, cases where Euclidean geometry may not be the
best option: Certain graphs may be known to have a structure which
would be best realized in a different geometry, such as on the surface
of a sphere or on a torus. In particular, 3D mesh data can be
parameterized on the sphere for texture mapping or graphs of genus one
can be embedded on a torus without crossings.  Furthermore, it has
also been noted that certain non- Euclidean geometries, specifically
hyperbolic geometry, have properties which are particularly well
suited to the layout and visualization of large classes of
graphs~\cite{EVL-1995-206,Munzner+1997a}. With this in mind, Euclidean force-directed algorithms have been extended to Riemannian spaces~\cite{kw-nese-05}.


\section{Spring Systems and Electrical Forces}
\label{fd:sec:fr}

The 1984 algorithm of Eades~\cite{Eades+1984a} targets graphs with up
to 30 vertices and uses a mechanical model to produce ``aesthetically
pleasing'' 2D layouts for plotters and CRT screens. The algorithm is
succinctly summarized as follows:

\begin{quote}
{\em To embed a graph we replace the vertices by steel rings and
replace each edge with a spring to form a mechanical system. The
vertices are placed in some initial layout and let go so that the
spring forces on the rings move the system to a minimal energy
state. Two practical adjustments are made to this idea: firstly,
logarithmic strength springs are used; that is, the force exerted by a
spring is: $c_1* \log(d/c_2),$ where $d$ is the length of the
spring, and $c_1$ and $c_2$ are constants. Experience shows that
Hookes Law (linear) springs are too strong when the vertices are far
apart; the logarithmic force solves this problem. Note that the springs
exert no force when $d=c_2$. Secondly, we make nonadjacent vertices
repel each other. An inverse square law force, $c_3/\sqrt{d},$ where
$c_3$ is constant and $d$ is the distance between the vertices, is
suitable.}
\end{quote}

\begin{figure}[t]
\begin{center}
\includegraphics[width=.31\textwidth]{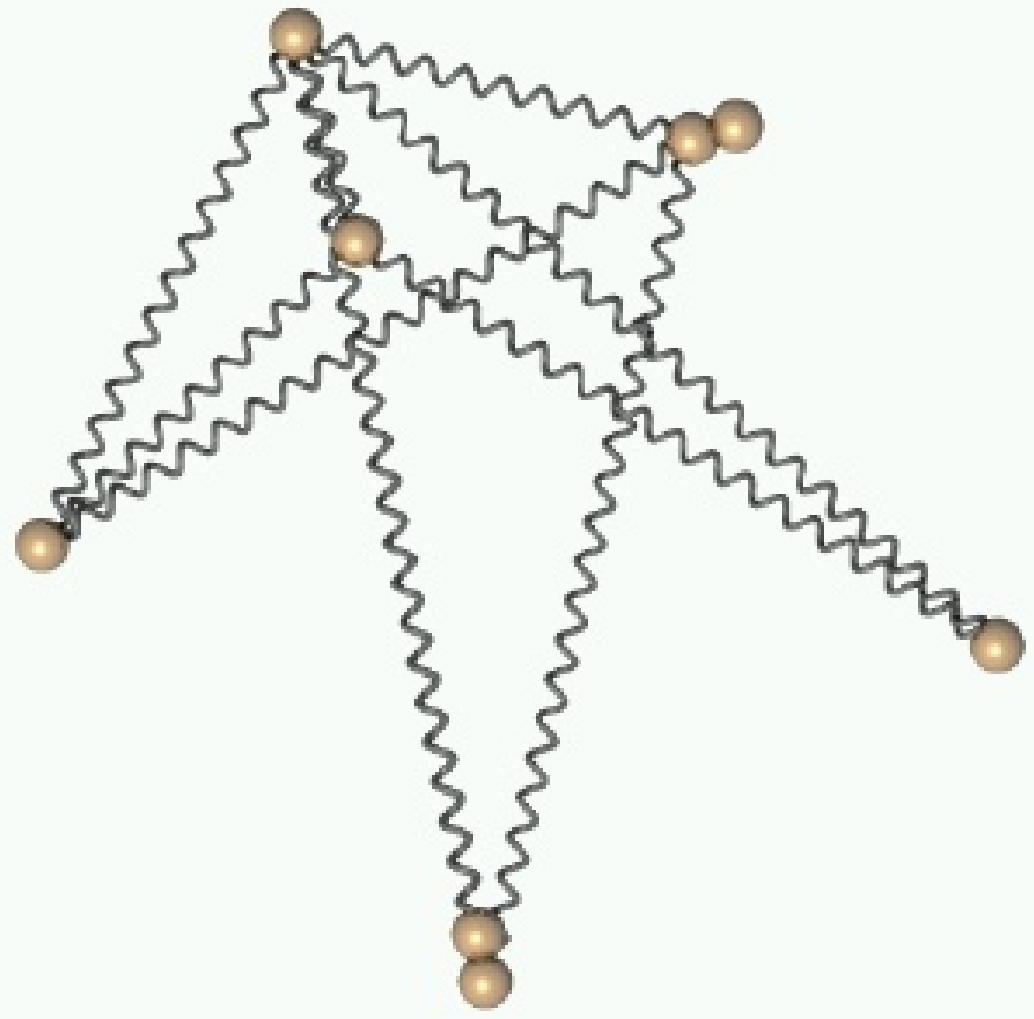}
\includegraphics[width=.31\textwidth]{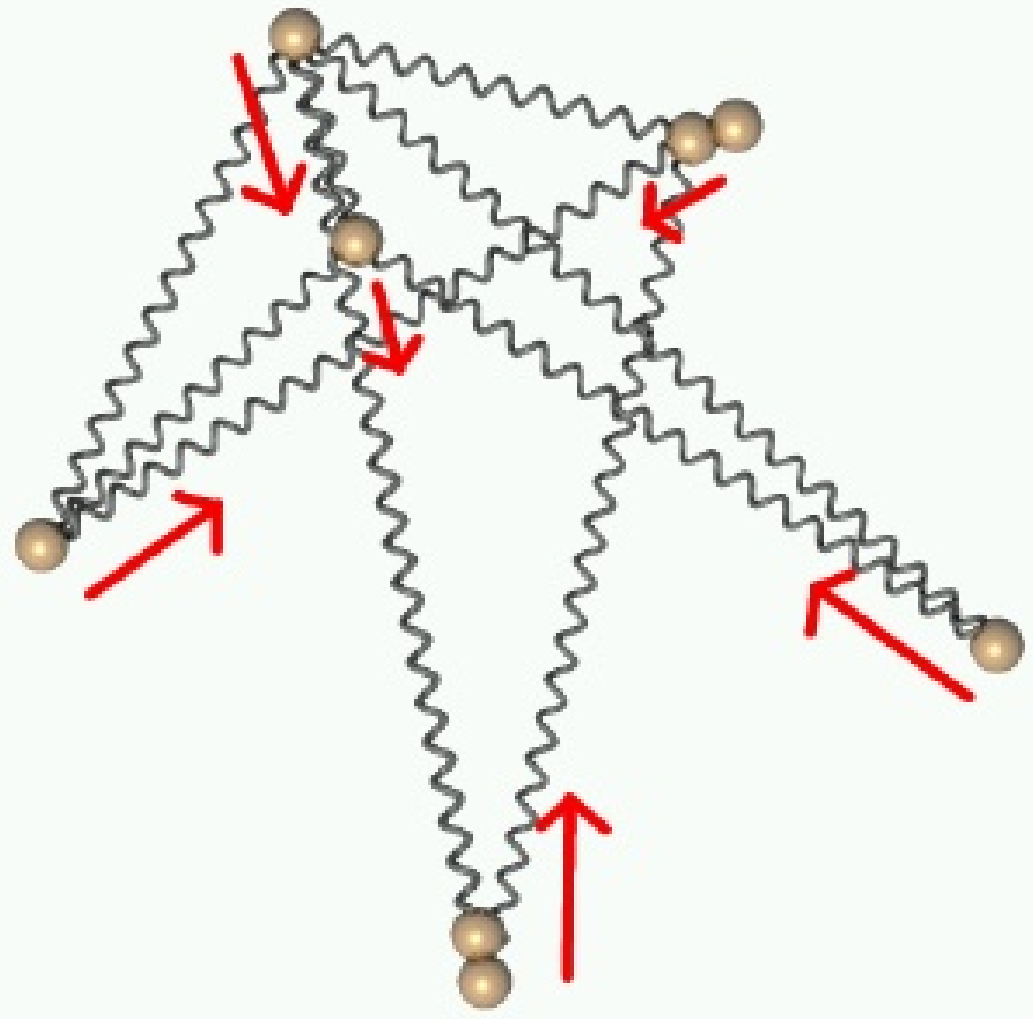}
\includegraphics[width=.34\textwidth]{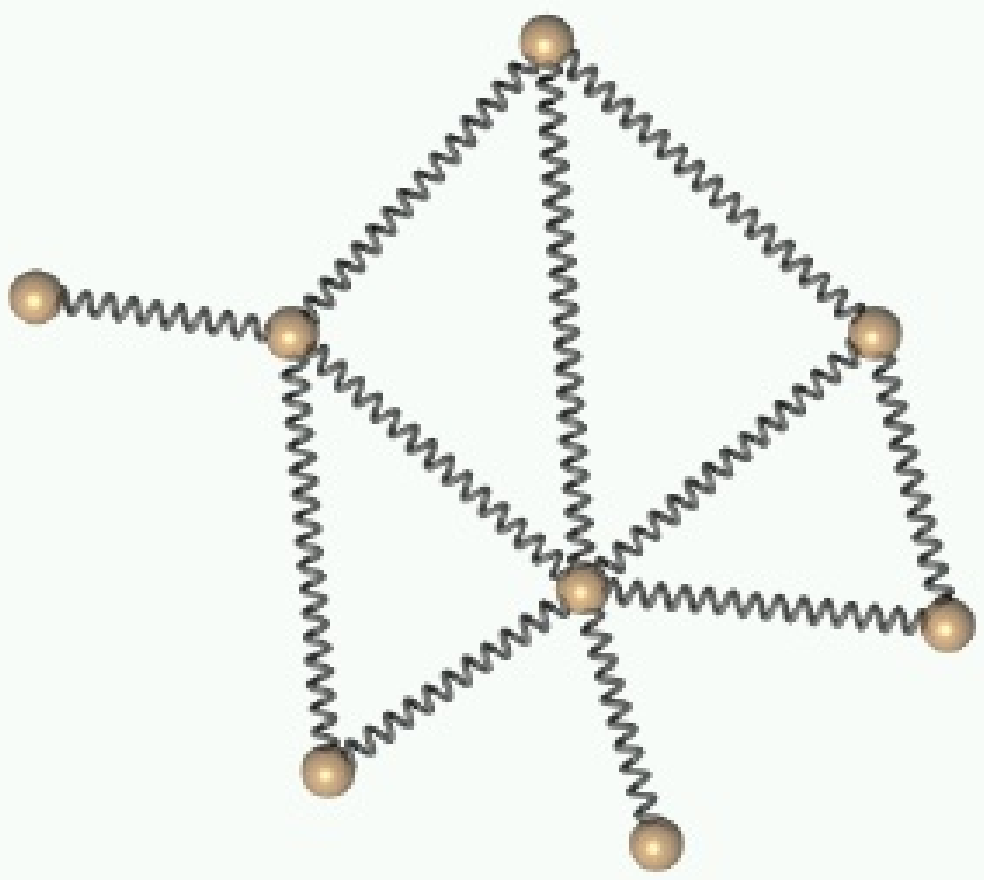}
\end{center}
\caption{\small\sf Illustration of a generic spring embedder: starting from random positions, treat the graph as spring system and look for a stable configuration~\cite{gk-grip-00}.
}
\label{fd:fig:spring}
\end{figure}

\begin{algorithm}
\KwIn{Graph $G$}
\KwOut{Straight-line drawing of $G$}
{\bf Initialize Positions:} place vertices of $G$ in random locations\;
\For{$i=1$ \KwTo $M$}{
calculate the force acting on each vertex\;
move the vertex $c_4 * (\mbox{force on vertex})$\;
}
draw a filled circle for each vertex\;
draw a straight-line segment for each edge\;
\caption{SPRING \label{alg:Spr}}
\end{algorithm}

This excellent description encapsulates the essence of spring
algorithms and their natural simplicity, elegance, and conceptual
intuitiveness; see Fig.~\ref{fd:fig:spring}. The goals behind ``aesthetically pleasing'' layouts
were initially captured by the two criteria: ``all the edge lengths
ought to be the same, and the layout should display as much symmetry
as possible.'' The resulting {\tt SPRING} algorithm of Eades is shown in Algorithm~\ref{alg:Spr}. The values $c_1=2$, $c_2=1$, $c_3=1$, $c_4=0.1$, and $M=100$ were used in the original paper. 

The 1991 algorithm of Fruchterman and Reingold added ``even vertex
distribution'' to the earlier two criteria and treats vertices in the
graph as ``atomic particles or celestial bodies, exerting attractive
and repulsive forces from one another.'' The attractive and repulsive
forces are redefined to $$f_a(d)=d^2/k,\mbox{\hspace{2cm}}
f_r(d)=-k^2/d,$$
in terms of the distance $d$ between two vertices and the optimal distance between vertices $k$ defined as $$k=C\sqrt{\frac{area}{number\ of\ vertices}}.$$

This algorithm is similar to that of Eades in that both algorithms
compute attractive forces between adjacent vertices and repulsive
forces between all pairs of vertices. The algorithm of Fruchterman and
Reingold adds the notion of ``temperature'' which could be used as
follows: ``the temperature could start at an initial value (say one
tenth the width of the frame) and decay to $0$ in an inverse linear
fashion.'' The temperature controls the displacement of vertices so
that as the layout becomes better, the adjustments become smaller. The
use of temperature here is a special case of a general technique
called simulated annealing, whose use in force-directed algorithms is
discussed later in this chapter. The pseudo-code for the Fruchterman
and Reingold provides further insight into the workings of a
spring-embedder; see Algorithm~\ref{alg:FR}.

\begin{algorithm}
$area\gets W * L$ \tcc*[r]{frame: width $W$ and length $L$}
initialize $G=(V,E)$ \tcc*[r]{place vertices at random}
$k\gets \sqrt{area/|V|}$ \tcc*[r]{compute optimal pairwise distance}
{\bf function} $f_r(x) = k^2/x$ \tcc*[r]{compute repulsive force}
\For{$i=1$ \KwTo $iterations$}{
\ForEach{$v \in V$}{
$v.disp:=0;$ \tcc*[r]{initialize displacement vector}
\For{$u \in V$}{
\If{$(u\neq v)$}{
$\Delta\gets v.pos-u.pos$ \tcc*[r]{distance between $u$ and $v$}
$v.disp\gets v.disp+(\Delta/|\Delta|)*f_r(|\Delta|)$ \tcc*[r]{displacement}
}
}
}
{\bf function} $f_a(x) = x^2/k$ \tcc*[r]{compute attractive force}
\ForEach{$e \in E$}{
$\Delta\gets e.v.pos-e.u.pos$ \tcc*[r]{$e$ is ordered vertex pair $.v$ and $.u$}
$e.v.disp\gets e.v.disp-(\Delta/|\Delta|)* f_a(|\Delta|)$\;
$e.u.disp\gets e.u.disp+(\Delta/|\Delta|)* f_a(|\Delta|)$\;
}
\ForEach{$v\in V$}{
\tcc*[f]{limit max displacement to frame; use temp.~$t$ to scale}
$v.pos \gets v.pos+(v.disp/|v.disp|)*\min(v.disp,t)$\;
$v.pos.x\gets \min(W/2, \max(-W/2, v.pos.x))$\;
$v.pos.y\gets \min(L/2, \max(-L/2, v.pos.y))$\;
}
$t\gets cool(t)$ \tcc*[r]{reduce temperature for next iteration}

}
\caption{Fruchterman-Reingold\label{alg:FR}}
\end{algorithm}

Each iteration the basic algorithm computes $O(|E|)$ attractive forces
and $O(|V|^2)$ repulsive forces. To reduce the quadratic complexity of
the repulsive forces, Fruchterman and Reingold suggest using a grid
variant of their basic algorithm, where the repulsive forces between
distant vertices are ignored. For sparse graphs, and with uniform
distribution of the vertices, this method allows a $O(|V|)$ time
approximation to the repulsive forces calculation. This approach can
be though of as a special case of the multi-pole technique introduced
in $n$-body simulations~\cite{ref:Greengard:1988a} whose use in
force-directed algorithms will be further discussed later in this
chapter.

As in the paper by Eades~\cite{Eades+1984a} the graphs considered by
Fruchterman and Reingold are small graphs with less than 40
vertices. The number of iterations of the main loop is also similar at
50.


\section{The Barycentric Method}
\label{fd:sec:bar}

Historically, Tutte's 1963 barycentric method is the
first ``force-directed'' algorithm for obtaining a straight-line,
crossings free drawing for a given 3-connected planar graph~\cite{t-hdg-63}. Unlike
almost all other force-directed methods, Tutte's guarantees that the
resulting drawing is crossings-free; moreover, all faces of the
drawing are convex.

The idea behind Tutte's algorithm is that if a face of the planar
graph is fixed in the plane, then suitable positions for the remaining
vertices can be found by solving a system of linear equations, where
each vertex position is represented as a convex combination of the
positions of its neighbors. This can be considered a force-directed
method as summarized in Di Battista {\em et al.}~\cite{dett-gd}; see Algorithm~\ref{alg:BC}.

\begin{algorithm}
\KwIn{$G=(V,E)$ with $V=V_0\cap V_1$, with fixed vertices $V_0$ and free vertices $V_1$; a strictly convex polygon $P$ with $|V_0|$ vertices.}
\KwOut{a position $p_v$ for each vertex of $V$, such that the fixed vertices form a convex polygon $P$.}
{\bf Initialize $V_0$:} place fixed vertices $u\in V_0$ at corners of $P$\;
{\bf Initialize $V_1$:} place free vertices $v\in V_1$ at the origin\;
\Repeat{$x_v$ and $y_v$ converge for all free vertices $v$}{
\ForEach{free vertex $v\in V_1$}{
$x_v\gets \frac{1}{deg(v)}\sum_{(u,v)\in E} x_u$\;
 $y_v\gets \frac{1}{deg(v)}\sum_{(u,v)\in E} y_u$\;
}
}
\caption{Barycenter-Draw\label{alg:BC}}
\end{algorithm}

In this model the force due to an edge $(u,v)$ is proportional to the
distance between vertices $u$ and $v$ and the springs have ideal
length of zero; there are no explicit repulsive forces. Thus the force
at a vertex $v$ is described by $$F(v)=\sum_{(u,v)\in E}(p_u - p_v),$$
where $p_u$ and $p_v$ are the positions of vertices $u$ and $v$. As
this function has a trivial minimum with all vertices placed in the
same location, the vertex set is partitioned into fixed and free
vertices. Setting the partial derivatives of the force function to
zero results in independent systems of linear equations for the
$x$-coordinate and for the $y$-coordinate.

The equations in the for-loop are linear and the number of equations
is equal to the number of the unknowns, which in turn is equal to the
number of free vertices. Solving these equations results in placing
each free vertex at the barycenter of its neighbors. The system of
equations can be solved using the Newton-Raphson method. Moreover, the
resulting solution is unique. 

One significant drawback of this approach is the resulting drawing
often has poor vertex resolution. In fact, for every $n>1$, there
exists a graph, such that the barycenter method computes a drawing
with exponential area~\cite{conf/gd/EadesG95}.

\section{Graph Theoretic Distances Approach}
\label{fd:sec:kk}

The 1989 algorithm of Kamada and Kawai~\cite{kk-adgug-89} introduced a
different way of thinking about ``good'' graph layouts. Whereas the
algorithms of Eades and Fruchterman-Reingold aim to keep adjacent
vertices close to each other while ensuring that vertices are not too
close to each other, Kamada and Kawai take graph theoretic approach:

\begin{quote}
{\em ``We regard the desirable geometric (Euclidean) distance between
two vertices in the drawing as the ``graph theoretic distance between
them in the corresponding graph.''}
\end{quote}

In this model, the ``perfect'' drawing of a graph would be one in
which the pair-wise geometric distances between the drawn vertices
match the graph theoretic pairwise distances, as computed by an
All-Pairs-Shortest-Path computation. As this goal cannot always be
achieved for arbitrary graphs in 2D or 3D Euclidean spaces, the
approach relies on setting up a spring system in such a way that
minimizing the energy of the system corresponds to minimizing the
difference between the geometric and graph distances. In this model
there are no separate attractive and repulsive forces between pairs of
vertices, but instead if a pair of vertices is (geometrically)
closer/farther than their corresponding graph distance the vertices
repel/attract each other. Let $d_{i,j}$ denote the shortest path
distance between vertex $i$ and vertex $j$ in the graph. Then
$l_{i,j}=L\times d_{i,j}$ is the ideal length of a spring between
vertices $i$ and $j$, where $L$ is the desirable length of a single
edge in the display. Kamada and Kawai suggest that $L=L_0/\max_{i<j}
d_{i,j}$, where $L_0$ is the length of a side of the display area and
$\max_{i<j} d_{i,j}$ is the diameter of the graph, i.e., the distance
between the farthest pair of vertices. The strength of the spring
between vertices $i$ and $j$ is defined as $$k_{i,j}=K/d_{i,j}^2,$$
where $K$ is a constant. Treating the drawing problem as
localizing $|V|=n$ particles $p_1, p_2,\dots, p_n$ in 2D Euclidean
space, leads to the following overall energy function:
$$E=\sum_{i=1}^{n-1}\sum_{j=i+1}^{n}\frac{1}{2}k_{i,j}(|p_i-p_j|-l_{i,j})^2.$$

The coordinates of a particle $p_i$ in the 2D Euclidean plane are given by $x_i$ and $y_i$ which allows us to rewrite the energy function as follows:
$$E=\sum_{i=1}^{n-1}\sum_{j=i+1}^{n}\frac{1}{2}k_{i,j}\left( (x_i-x_j)^2+(y_i-y_j)^2+l_{i,j}^2-2l_{i,j}\sqrt{(x_i-y_i)^2+(y_i-y_j)^2} \right).$$

The goal of the algorithm is to find values for the variables that
minimize the energy function $E(x_1, x_2, \dots, x_n, y_1, y_2, \dots, y_n)$. In
particular, at a local minimum all the partial derivatives are equal
to zero, and which corresponds to solving $2n$ simultaneous non-linear
equations. Therefore, Kamada and Kawai compute a stable position one
particle $p_m$ at a time. Viewing $E$ as a function of only $x_m$ and
$y_m$ a minimum of $E$ can be computed using the Newton-Raphson
method. At each step of the algorithm the particle $p_m$ with the largest value of $\Delta_m$ is chosen, where
$$\Delta_m=\sqrt{\left( \frac{\partial E}{\partial x_m}\right)^2+ \left( \frac{\partial E}{\partial y_m}\right)^2}.$$
This leads to the algorithm Kamada-Kawai; see Algorithm~\ref{alg:KK}.

\begin{algorithm}
compute pairwise distances $d_{i,j}$ for $1\leq i\neq j \leq n$;\\
compute pairwise ideal lengths $l_{i,j}$ for $1\leq i\neq j \leq n$;\\
compute pairwise spring strength $k_{i,j}$ for $1\leq i\neq j \leq n$;\\
initialize particle positions $p_1, p_2, \dots, p_n$;\\
\While{$(max_i\Delta_i>\epsilon)$}{
let $p_m$ be the particle satisfying $\Delta_m=max_i\Delta_i$\;
\While{$(\Delta_m>\epsilon)$}{
compute $\delta x$ and $\delta y$ by solving the following system of equations:\\
 \hspace{.5cm} $\frac{\partial^2 E}{\partial x_m^2}(x_m^{(t)},y_m^{(t)})\delta x + \frac{\partial^2 E}{\partial x_m \partial y_m}(x_m^{(t)},y_m^{(t)})\delta y= - \frac{\partial E}{\partial x_m}(x_m^{(t)},y_m^{(t)})$;\\
\hspace{.5cm} $\frac{\partial^2 E}{\partial y_m \partial x_m}(x_m^{(t)},y_m^{(t)})\delta x + \frac{\partial^2 E}{\partial y_m^2}(x_m^{(t)},y_m^{(t)})\delta y= - \frac{\partial E}{\partial y_m}(x_m^{(t)},y_m^{(t)})$\\
 $x_m\gets x_m+\delta x$\;
 $y_m\gets y_m+\delta y$\;
}
}
\caption{Kamada-Kawai\label{alg:KK}}
\end{algorithm}

The algorithm of Kamada and Kawai is computationally expensive,
requiring an All-Pair-Shortest-Path computation which can be done in
$O(|V|^3)$time using the Floyd-Warshall algorithm or in $O(|V|^2 \log
|V| +|E||V|)$ using Johnson's algorithm;
see the All-Pairs-Shortest-Path chapter
in an algorithms textbook such as~\cite{clrs-ia-90}. Furthermore, the
algorithm requires $O(|V|^2)$ storage for the pairwise vertex
distances. Despite the higher time and space complexity, the algorithm
contributes a simple and intuitive definition of a ``good'' graph
layout: A graph layout is good if the geometric distances between
vertices closely correspond to the underlying graph distances.


\section{Further Spring Refinements}
\label{fd:sec:ar} 

Even before the 1984 algorithm of Eades, force-directed techniques
were used in the context of VLSI layouts in the 1960's and
1970's~\cite{fcw-accel-67,qb-fdcpp-79}. Yet, renewed interest in
force-directed graph layout algorithms brought forth many new ideas in
the 1990's. Frick {\em et al.}~\cite{flm-falau-95} add new
heuristics to the Fruchterman-Reingold approach. In particular,
oscillation and rotations are detected and dealt with using local
instead of global temperature. The following year Bru{\ss} and
Frick~\cite{bf-fi3dg-96} extended the approach to layouts directly in
3D Euclidean space. The algorithm of Cohen~\cite{ACMTOCHI::Cohen1997}
introduced the notion of an incremental layout, a precursor of the
multi-scale methods described in Section~\ref{fd:sec:large}.

The 1997 algorithm of Davidson and Harel~\cite{dh-dgnus-96} adds
additional constraints to the traditional force-directed approach in
explicitly aiming to minimize the number of edge-crossings and keeping
vertices from getting too close to non-adjacent edges. The algorithm
uses the simulated annealing technique developed for large
combinatorial optimization~\cite{kirkpatrick83optimization}. Simulated
annealing is motivated by the physical process of cooling molten
materials. When molten steel is cooled too quickly it cracks and forms
bubbles making it brittle. For better results, the steel must be
cooled slowly and evenly and this process is known as annealing in
metallurgy. With regard to force-directed algorithms, this process is
simulated to find local minima of the energy function. 

Genetic algorithms for force-directed placement have also been
considered. Genetic algorithms are a commonly used search technique
for finding approximate solutions to optimization and search
problems. The technique is inspired by evolutionary biology in general
and by inheritance, mutation, natural selection, and recombination (or
crossover), in particular; see the survey by Vose~\cite{v-sga-99}.  In
the context of force-directed techniques for graph drawing, the
genetic algorithms approach was introduced in 1991 by Kosak {\em et al.}~\cite{kosak91}. Other notable approaches in the direction include that of Branke {\em et al.}~\cite{Branke97}.

In the context of graph clustering, the {\em LinLog} model introduces
an alternative energy model~\cite{Noack07}. Traditional energy models enforce small and uniform edge
lengths, which often prevents the separation of nodes in different
clusters. As a side effect, they tend to group nodes with large degree
in the center of the layout, where their
distance to the remaining nodes is relatively small. The node-repulsion LinLog and edge-
repulsion LinLog models, group nodes according to two
well-known clustering criteria: the density of the cut~\cite{lr-amf-88} and the normalized cut~\cite{sm-nc-00}. 

\section{Large Graphs}
\label{fd:sec:large}

The first force-directed algorithms to produce good layouts for graphs
with over 1000 vertices is the 1999 algorithm of Hadany and
Harel~\cite{hh-msadg-99}. They introduced the multi-scale technique as
a way to deal with large graphs and in the following year four related
but independent force-directed algorithms for large graphs were
presented at the Annual Symposium on Graph Drawing. We begin with
Hadany and Harel's description on the multi-scale method:

\begin{quote}
{\em
A natural strategy for drawing a graph nicely is to first consider an abstraction, disregarding some of the graph's fine details. This abstraction is then drawn, yielding a ``rough'' layout in which only the general structure is revealed. Then the details are added and the layout is corrected. To employ such a strategy is it crucial that the abstraction retain essential features of the graph. Thus, one has to define the notion of coarse-scale representations of a graph, in which the combinatorial structure is significantly simplified but features important for visualization are well preserved. The drawing process will then ``travel'' between these representations, and introduce multi-scale corrections. Assuming we have already defined the multiple levels of coarsening, the general structure of our strategy is as follows:
\begin{enumerate}
\item Perform {\em fine-scale} relocations of vertices that yield a locally organized configuration.
\item Perform {\em coarse-scale} relocations (through local relocations in the coarse representations), correcting global disorders not found in stage 1.
\item Perform {\em fine-scale} relocations that correct local disorders introduced by stage 2.
\end{enumerate}
}\end{quote}

Hadany and Harel suggest computing the sequence of graphs by using
edge contractions so as to preserve certain properties of the
graph. In particular, the goal is to preserve three topological
properties: cluster size, vertex degrees, and homotopy. For the
coarse-scale relocations, the energy function for each graph in the
sequence is that of Kamada and Kawai (the pairwise graph distances are
compared to the geometric distances in the current layout). For the
fine-scale relocations, the authors suggest using force-directed
calculations as those of Eades~\cite{Eades+1984a},
Fruchterman-Reingold~\cite{fr-gdfdp-91}, or
Kamada-Kawai~\cite{kk-adgug-89}. While the asymptotic complexity of
this algorithm is similar to that of the Kamada-Kawai algorithm, the
multi-scale approach leads to good layouts for much larger graphs in
reasonable time. 

The 2000 algorithm of Harel and Koren~\cite{hk-fmsmd-j-02} took
force-directed algorithms to graphs with 15,000 vertices. This
algorithm is similar to the algorithm of Hadany and Harel, yet uses a
simpler coarsening process based on a $k$-centers approximation, and a
faster fine-scale beautification. Given a graph $G=(V,E)$, the
$k$-centers problem asks to find a subset of the vertex set
$V'\subseteq V$ of size $k$, so as to minimize the maximum distance
from a vertex to $V'$: $\min{u\in V}\max_{u\in V, v\in V'} dist(u,v)$.
While $k$-centers is an NP-hard problem, Harel and Koren use a
straightforward and efficient 2-approximation algorithm that relies on
Breadth-First Search~\cite{Hochbaum-96}. The fine-scale vertex
relocations are done using the Kamada-Kawai approach; see Algorithm~\ref{alg:HK}.

\begin{algorithm}
\KwIn{graph $G=(V,E)$}
\KwOut{Find $L$, a nice layout of $G$}
$Iterations=4$ \tcc*[r]{number of iterations in local beautification}
$Ratio=3$ \tcc*[r]{ratio of vertex sets in consecutive levels}
$Rad=7$ \tcc*[r]{radius of local neighborhood}
$MinSize=10$ \tcc*[r]{size of coarsest graph}
Compute all-pairs shortest path lengths $d_{i,j}$\;
Initialize layout $L$ by placing vertices at random\;
$k\gets MinSize$\;
\While{$k\leq |V|$}{
$C \gets ${\bf K-Centers}$(G(V,E), k)$\;
$radius = \max_{v\in C} \min_{u\in C}\{d_{vu}\}*Rad$\;
{\bf LocalLayout}$(d_{C\times C}, L(C),radius,Iterations)$\;
\ForEach{ $v \in V$}{
$L(v) \in L(center(v)) + rand$\;
}
$k \gets k · Ratio$\\
}
\Return{$L$}\;

{\bf K-Centers}$(G(V,E), k)$\;
\KwIn{Graph $G=(V,E)$ and constant $k$}
\KwOut{Compute set $S \subseteq V$ of size $k$, s.t.~$\max_{v\in V}\min_{s\in S}\{d_{sv}\}$ is minimized}
$S \gets \{v\}$ for some arbitrary $v \in V$\;
\For{$i = 2$ to $k$}{
find vertex $u$ farthest away from $S$\;
(i.e., such that $\min_{s\in S}\{d_{us}\} \geq  \min_{s\in S}\{d_{ws}\}, \forall w \in V$ )\;
$S \gets S \cup \{u\}$\;
}
\Return{$S$}\;

{\bf LocalLayout}$(d_{V\times V} , L, k, Iterations)$\;
\KwIn{APSP matrix $d_{V\times V}$, layout $L$, constants $k$ and $Iterations$}
\KwOut{Compute locally nice layout $L$ by beautifying $k$-neighborhoods}
\For{$i = 1$ \KwTo $Iterations*|V|$}{
Choose the vertex $v$ with the maximal $\Delta_v^k$\;
Compute $\delta_v^k$ as in Kamada-Kawai\;
$L(v) \gets L(v) + (\delta_v^k(x),\delta_v^k(y))$\;
}
\caption{Harel and Koren\label{alg:HK}}
\end{algorithm}

The 2000 algorithm of Gajer {\em et al.}~\cite{ggk-afmda-00j} is also a multi-scale force-directed
algorithm but introduces several ideas to the realm of multi-scale
force directed algorithms for large graphs. Most importantly, this
approach avoids the quadratic space and time complexity of previous
force-directed approaches with the help of a simpler coarsening
strategy. Instead of computing a series of coarser graphs from the
given large graph $G=(V,E)$, Gajer {\em et al.} produce a vertex
filtration ${\cal V}:\ V_0 \supset V_1 \supset \ldots \supset V_k
\supset
\emptyset$, where $V_0=V(G)$ is the original vertex set of the given
graph $G$. By restricting the number of vertices considered in
relocating any particular vertex in the filtration and ensuring that
the filtration has $O(\log |V|)$ levels an overall running time of
$O(|V|\log^2 |V|)$ is achieved. In particular, filtrations based on 
maximal independent sets are considered. $ V = V_0 \supset V_1 \supset \ldots \supset V_k \supset
\emptyset$, is a maximal independent set filtration of $G$ if $V_i$ is a maximal subset of $V_{i-1}$ for which the
graph distance between any pair of its elements is greater than or
equal to $2^i$; see Algorithm~\ref{alg:main}.

\begin{algorithm}
\SetKw{KwDto}{downto}
\KwIn{Graph $G(V,E)$}
\KwOut{Straight-line drawing of $G$}
create a vertex filtration ${\cal V}:\ V_0 \supset V_1 \supset \ldots \supset V_k \supset \emptyset$\;
\For{i=k \emph{\KwDto}0}{
  \ForEach{$v \in V_i - V_{i+1}$}{
    find vertex neighborhood $N_{i}(v), N_{i-1}(v), \ldots, N_0(v)$\;
    find initial position $pos[v]$ of $v$\;
    }
    \For{j=0 \emph{\KwTo}$rounds$}{
      \ForEach{$v \in V_i$}{
        compute local temperature $heat[v]$\;
        $disp[v] \gets heat[v]\cdot{\overrightarrow{F}\hspace{-.1cm}_{N_i}(v)}$\;
        }
      \ForEach{$v \in V_i$}{
          $pos[v] \gets pos[v] + disp[v]$\;
          }
     }
}
add all edges $e \in E$\;
\caption{GRIP \label{alg:main}}
\end{algorithm}

\begin{figure}[t]
\begin{center}
\includegraphics[width=.32\textwidth]{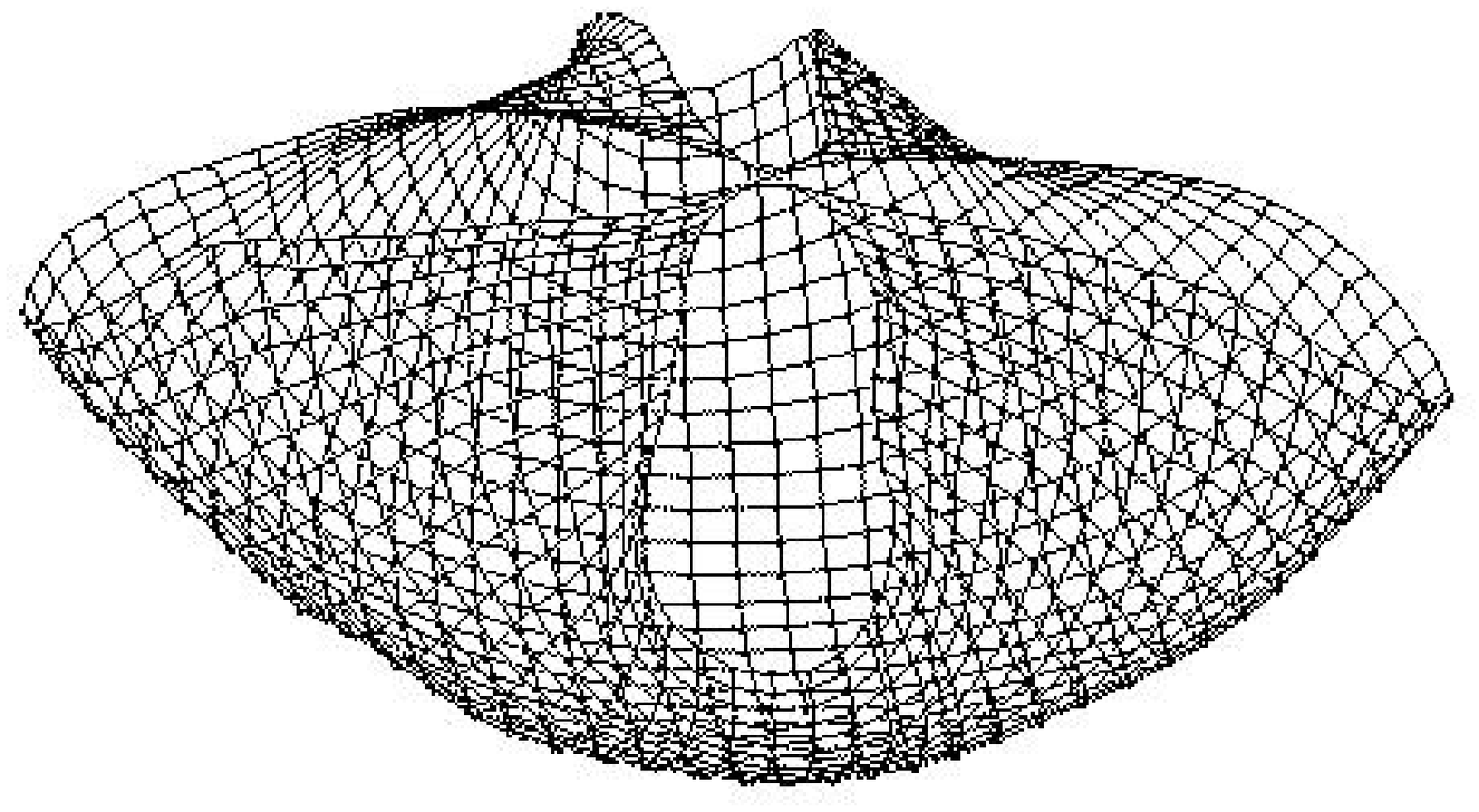}
\includegraphics[width=.32\textwidth]{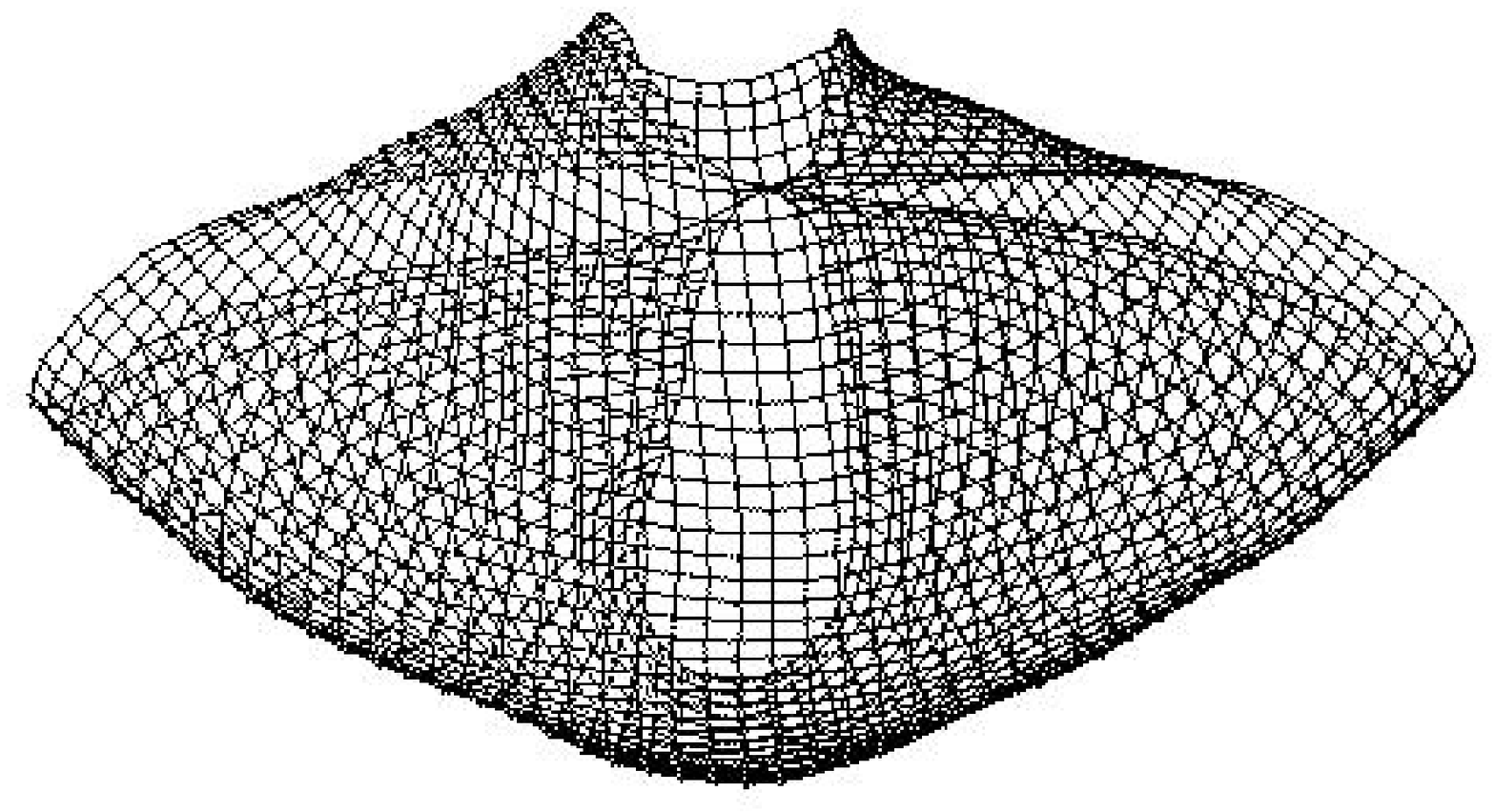}
\includegraphics[width=.32\textwidth]{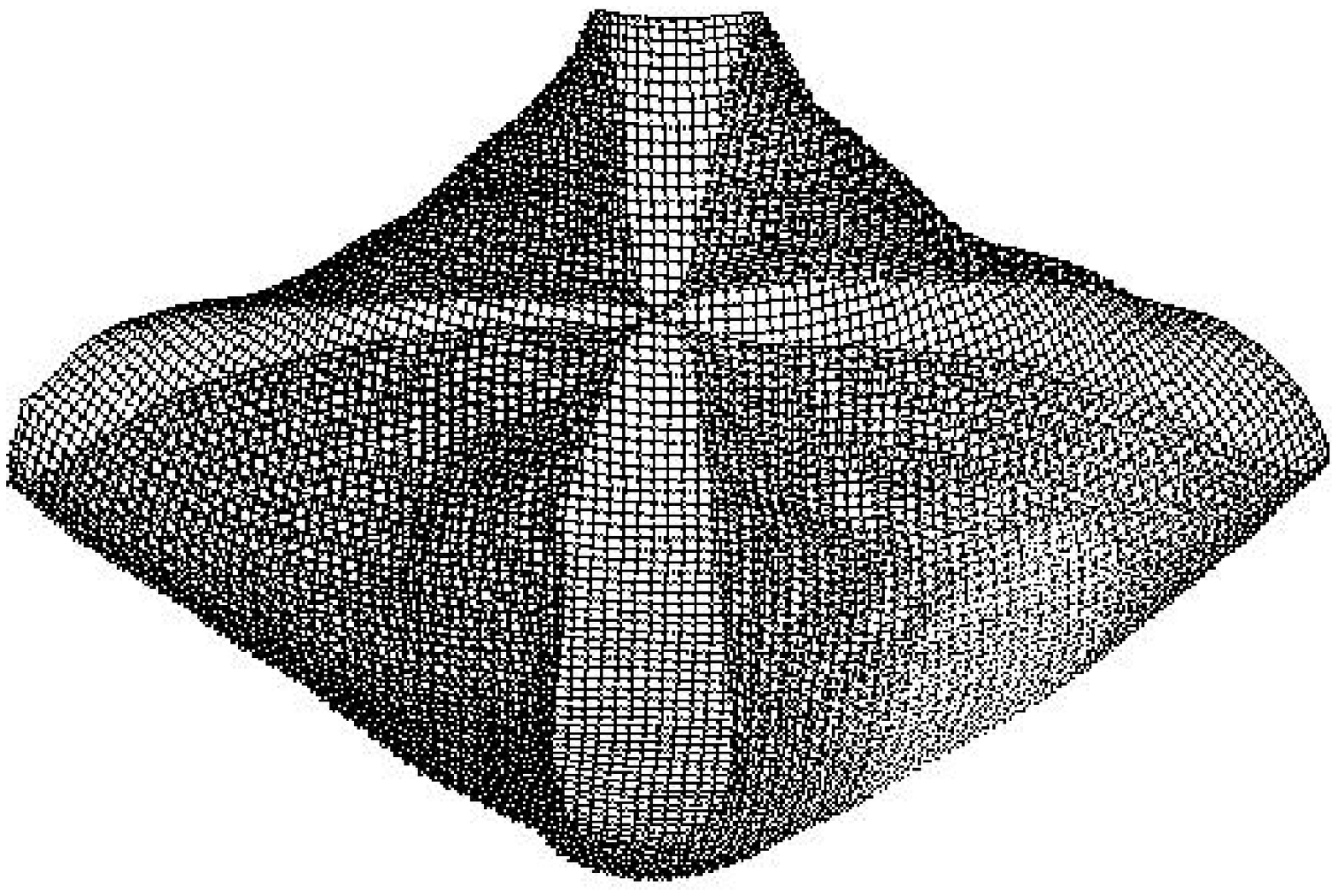}\\
\includegraphics[width=.3\textwidth]{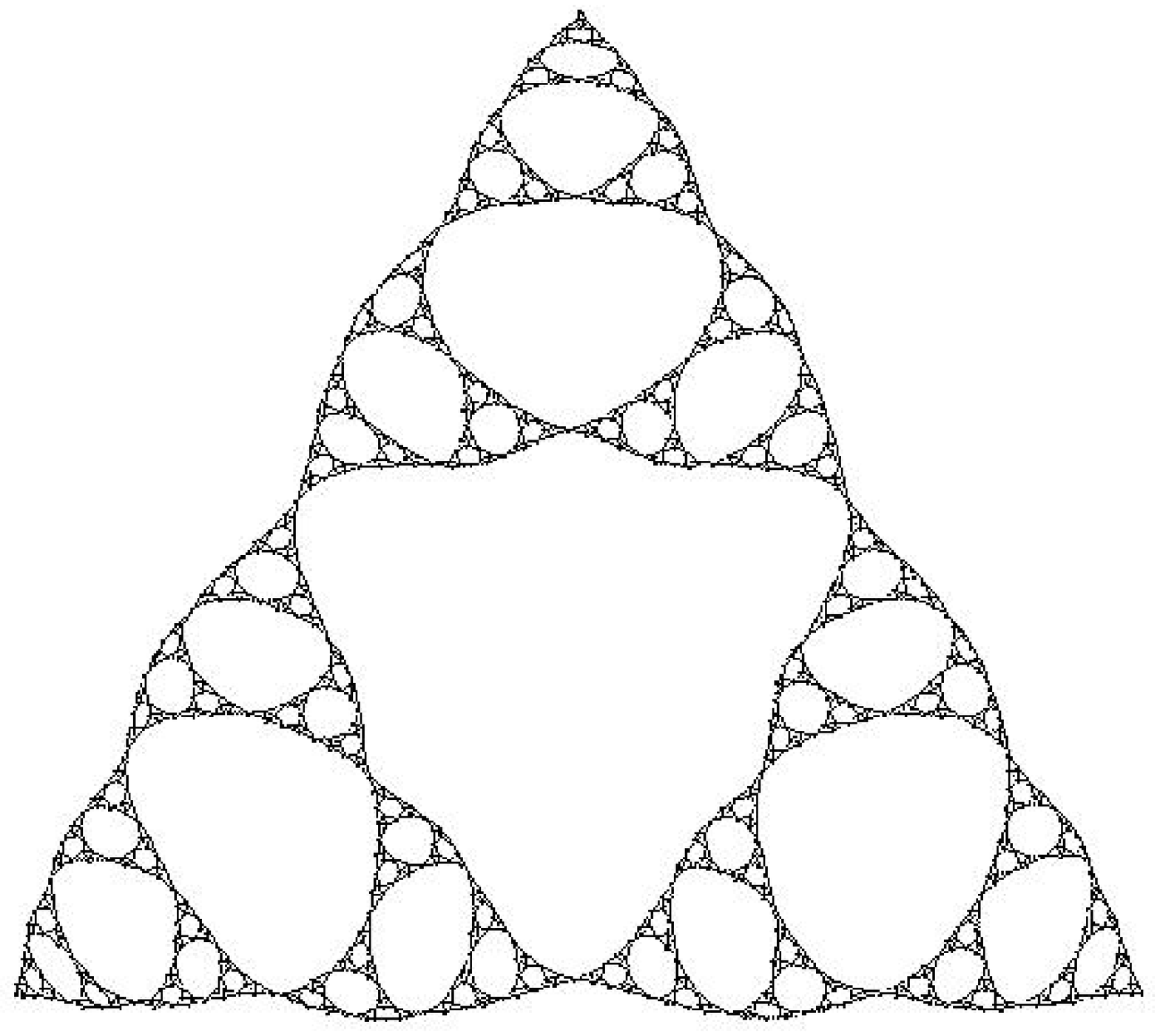}
\rotatebox{-4}{\includegraphics[width=.34\textwidth]{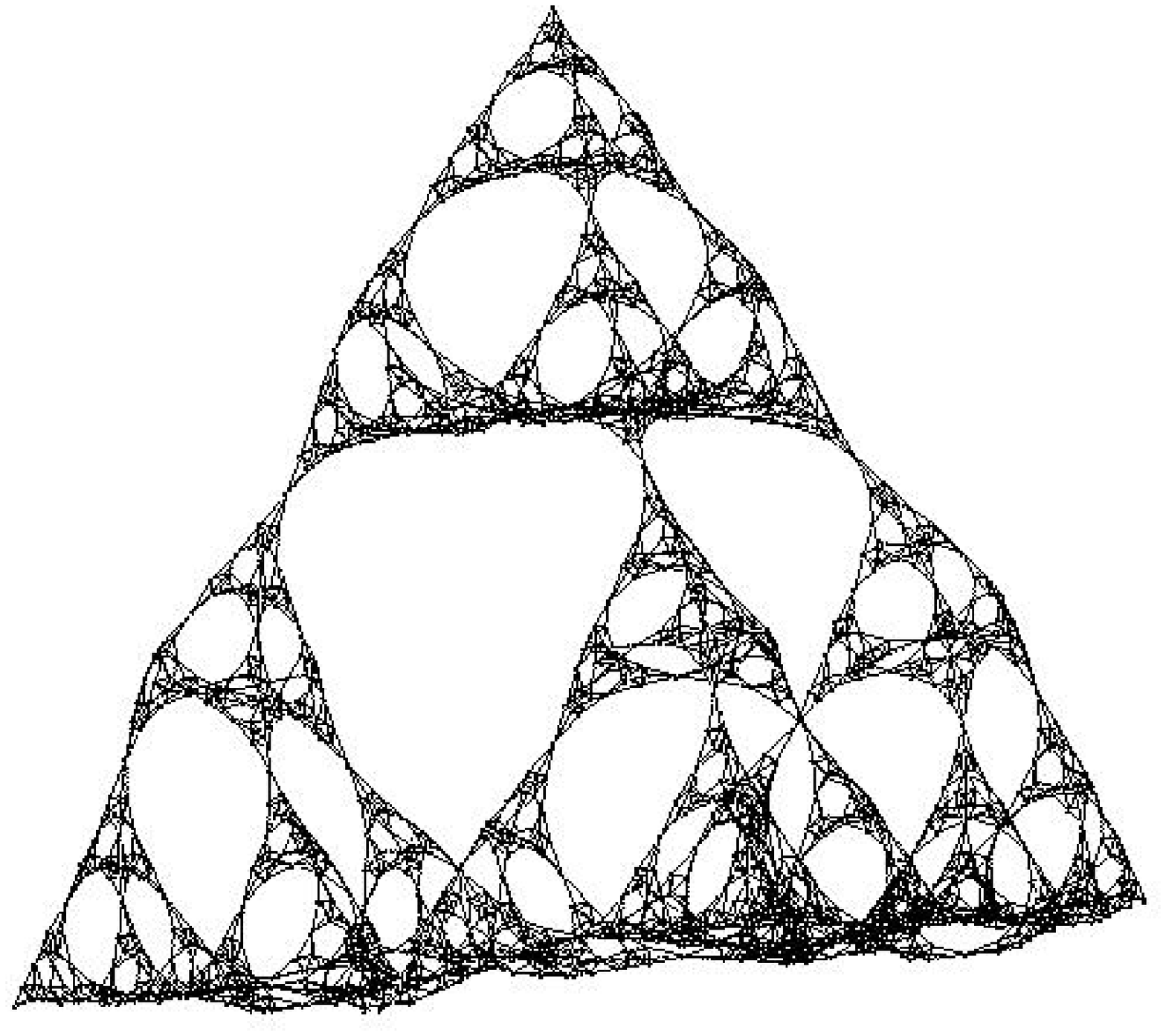}}
\includegraphics[width=.32\textwidth]{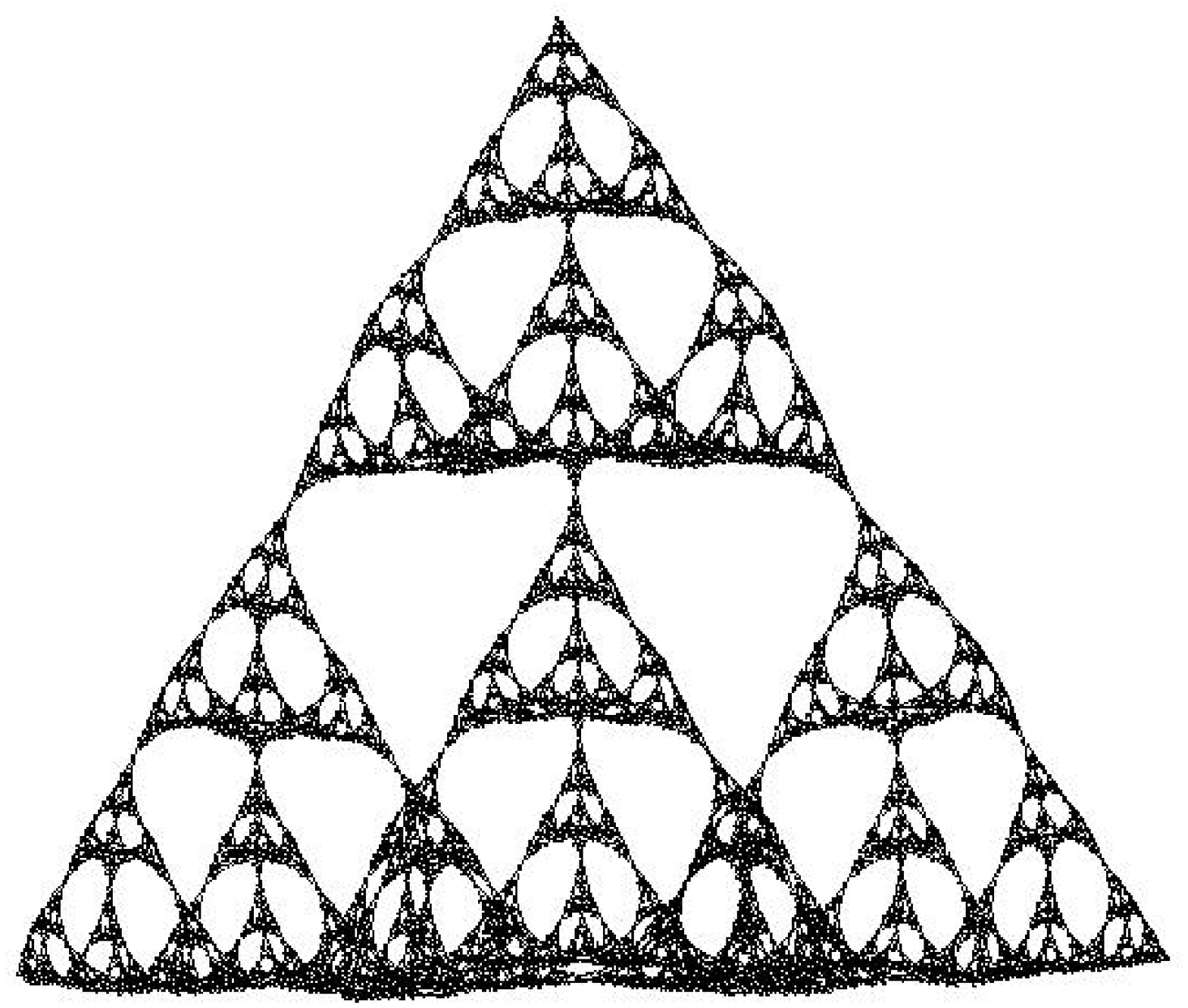}
\end{center}
\caption{\small\sf Drawings from GRIP. First row: knotted
  meshes of 1600, 2500, and 10000 vertices. Second row: Sierpinski graphs, 2D of order 7
(1,095 vertices), 3D of order 6 (2,050 vertices), 3D of order 7 (8,194 vertices)~\cite{gk-grip-00}.}
  \label{fd:fig:grip}
\end{figure}
The {\tt GRIP} system~\cite{gk-grip-00} builds on the filtration and neighborhood calculations
of~\cite{ggk-afmda-00j}. It introduces the idea of realizing the graph in
high-dimensional Euclidean space and obtaining 2D or 3D projections at
end. The algorithm also relies on intelligent initial placement of
vertices based on graph theoretic distances, rather than random
initial placement. Finally, the notion of cooling is re-introduced in
the context of multi-scale force-directed algorithms.  The {\tt GRIP}
system produces high-quality layouts, as illustrated in Fig~\ref{fd:fig:grip}.

Another 2000 multilevel algorithm is that
of Walshaw~\cite{w-mafdgd-j-03}. Instead of relying on the
Kamada-Kawai type force interactions, this algorithm extends the grid
variant of Fruchterman-Reingold to a  multilevel algorithm. The
coarsening step is based on repeatedly collapsing maximally
independent sets of edges, and the fine-scale refinements are based on
Fruchterman-Reingold force calculations. This results in $O(|V|^2)$ running time;  see Algorithm~\ref{alg:Wal}.

\begin{algorithm}
	   {\bf function} $f_g(x,w)= -Cwk^2/x$ \tcc*[r]{global repulsive force}
           {\bf function} $f_l(x,d,w)= \{(x-k)/d\} -f_g(x,w)$ \tcc*[r]{local spring force} 
	   $t\gets t_0$	\tcc*[r]{initial temperature}
	   $Posn \gets NewPosn$\;
           \While{$(converged\neq 1)$}{
	   $converged\gets 1$\;
           \lFor{$v\in V$}{ $OldPosn[v] \gets NewPosn[v]$}\;
	   \ForEach{$v\in V$}{
             $D\gets 0$ \tcc*[r]{initialize vector of displacements $D$ of $v$} 

	   \ForEach{$u \in V, u\neq v$}{
	   $\Delta\gets Posn[u]-Posn[v]$ \tcc*[r]{calculate global forces}
	   $D \gets D+(\Delta/|Delta|)*f_g(|\Delta|, |u|)$\;
	   }
	   \ForEach{$u \in \Gamma(v)$}{
	   $\Delta\gets Posn[u]-Posn[v]$ \tcc*[r]{calculate local forces}
	   $D\gets D+(\Delta/|Delta|)*f_l(|\Delta|, |\Gamma(v)|,|u|)$\;
	   }
	   $NewPosn[v]=NewPosn[v]+(D/|D|)*\min(t,|D|)$ 	 \tcc*[r]{move $v$}
	   $\Delta:=NewPosn[v]-OldPosn[v]$\;
	   \If{$(|\Delta|>k\times tol$)} {$converged\gets 0$}
	   }
	   $t\gets cool(t)$ \tcc*[r]{scale temperature to reduce maximum movement}

	   }
\caption{Walshaw \label{alg:Wal}}
\end{algorithm}

The fourth 2000 multilevel force-directed algorithm is due to Quigley
and Eades~\cite{qe-fade-00}. This algorithm relies on the Barnes-Hut
$n$-body simulation method~\cite{Barnes86a} and reduces repulsive
force calculations to $O(|V| \log |V|)$ time instead of the usual
$O(|V|^2)$.  Similarly, the algorithm of Hu~\cite{Hu05} combines the multilevel
approach with the $n$-bosy simulation method, and is implemented in
the {\tt sfdp} drawing engine of GraphViz~\cite{graphviz}.

One possible drawback to this approach is that the running
time depends on the distribution of the vertices. Hachul and
J{\"u}nger~\cite{hj-dlg-04} address this problem in their 2004 multilevel
algorithm. 


\section{Stress Majorization}
\label{fd:sec:stress}

Methods that exploit fast algebraic operations offer another practical
way to deal with large graphs. {\em Stress
  minimization} has been proposed and implemented in the more general
setting of multidimensional scaling (MDS)~\cite{oldKruskalMDS}. The
function describing the stress is similar to the layout energy function of Kamada-Kawai from
Section~\ref{fd:sec:kk}:
$$E=\sum_{i=1}^{n-1}\sum_{j=i+1}^{n}\frac{1}{2}k_{i,j}(|p_i-p_j|-l_{i,j})^2,$$
but here $k_{i,j}$=1 and $l_{i,j}=d_{i,j}$ is simply the graph
theoretic distance. In their paper on graph drawing by stress
minimization Gansner {\em et al.}~\cite{gkn-gdsm-04} point out that this
particular formulation of the energy of the layout, or {\em stress
  function} has been already used to draw graphs as early as in
1980~\cite{ks-dnd-80}. What makes this particular stress function relevant to drawing large
graphs is that it can be optimized better that
with the local Newton-Raphson method or with gradient descent. Specifically, this stress function can
be globally minimized via {\em majorization}. That is, unlike the
energy function of Kamada-Kawai, the classical MDS stress function can
be optimized via majorization which is guaranteed to converge.

The {\em strain} model, or classical scaling, is related to the stress
model. In this setting a solution can be obtained via an
eigen-decomposition of the adjacency matrix. 
Solving the full stress or strain model still requires computing all pairs
shortest paths. Significant savings can be gained if we instead
compute a good approximation. In PivotMDS Brandes and
Pich~\cite{bp-pmds-06} show that replacing the all-pairs-shortest path
computation with a distance calculations from a few vertices in the
graph is often sufficient, especially if combined with a solution to a
sparse stress model. 

When not all nodes are free to move, {\em constrained
stress majorization} can be used to support additional constraints by,
and treating the majorizing functions as a quadratic program~\cite{DBLP:journals/dm/DwyerKM09}.
Planar graphs are of particular interest in graph drawing, and often
force-directed graph drawing algorithms are used to draw
them. While in theory any planar graph has a straight-line
crossings-free drawing in the plane, force-directed algorithms do not
guarantee such drawings. 

Modifications to the basic force-directed functionality, with the aim of improving the layout quality for planar graphs, have also been considered. Harel and Sardas~\cite{harel98algorithm} improve an earlier simulated annealing drawing algorithm by Davidson and Harel~\cite{dh-dgnus-96}. The main idea is to obtain an initial plane embedding and then apply simulated annealing while not introducing any crossings. Overall their method significantly improved the aesthetic quality of the initial planar layouts, but at the expense of a significant increase in running time of $O(n^3)$, making it practical only for small graphs.
PrEd~\cite{Bertault-pred-00} and {ImPrEd}~\cite{simonetto-11} are
force-directed algorithms that improve already created drawings of a graph.
PrEd extends the method of Fruchterman and Reingold~\cite{fr-gdfdp-91}
and can be used as a post-processing crossings-preserving
optimization. In particular, PrEd takes some straight-line drawing as
input and guarantees that no new edge crossings will be created (while
preserving existing crossings, if any are present in the input
drawing). Then the algorithm can be used to optimize a planar layout,
while preserving its planarity and its embedding, or to improve a
graph that has a meaningful initial set of edge crossings. To achieve
this result, PrEd adds a phase where the maximal movement of each node
is computed, and adds a repulsive force between (node, edge) pairs. 
The main aims of ImPrEd~\cite{simonetto-11} are to significantly
reduce the running time of PrEd, achieve high aesthetics even for
large and sparse graphs, and make the algorithm more stable and
reliable with respect to the input parameters. This is achieved via
improved spacing of the graph elements and an accelerated convergence
of the drawing to its final configuration.


\section{Non-Euclidean Approaches}
\label{fd:sec:nea}

Much of the work on non-Euclidean graph drawing has been done in
hyperbolic space which offers certain advantages over Euclidean space;
see Munzner~\cite{Munzner+1997a,Munzner:1996:VSW}.  For example, in
hyperbolic space it is possible to compute a layout for a complete
tree with both uniform edge lengths and uniform distribution of nodes.
Furthermore, some of the embeddings of hyperbolic space into Euclidean
space naturally provide a fish-eye view of the space, which is useful
for ``focus+context'' visualization, as shown by Lamping {\em et
al.}~\cite{EVL-1995-206}. From a visualization point of view,
spherical space offers a way to present a graph in a center-free and
periphery-free fashion. That is, in traditional drawings in
$\mathbb{R}^2$ there is an implicit assumption that nodes in the
center are important, while nodes on the periphery are less
important. This can be avoided in $\mathbb{S}^2$ space, where any part
of the graph can become the center of the layout. The early approaches for calculating the
layouts of graphs in hyperbolic space, however, are either restricted
by their nature to the layout of trees and tree-like graphs, or to
layouts on a lattice.

The hyperbolic tree layout algorithms function on the principle of
hyperbolic sphere packing, and operate by making each node of a tree,
starting with the root, the center of a sphere in hyperbolic space.
The children of this node are then given positions on the surface of
this sphere and the process recurses on these children.  By carefully
computing the radii
of these spheres
it is possible to create aesthetically pleasing layouts for the
given tree. 

 Although some applications calculate the layout of a
general graph using this method, the layout is calculated using a
spanning tree of the graph and the extra edges are then added in
without altering the layout~\cite{munzner-gd98}. This method works
well for tree-like and quasi-hierarchical graphs, or for graphs
where domain-specific knowledge provides a way to create a
meaningful spanning tree. However, for general graphs (e.g.,
bipartite or densely connected graphs) and without relying on
domain specific knowledge, the tree-based approach may result in
poor layouts.

Methods for generalizing Euclidean geometric algorithms to hyperbolic
space, although not directly related to graph drawing, have also been
studied. Recently, van Wijk and Nuij~\cite{wn-msv-04} proposed a
Poincar\'{e}'s half-plane projection to define a model for 2D viewing
and navigation. Eppstein~\cite{eppstein-msri-03} shows that many
algorithms which operate in Euclidean space can be extended to
hyperbolic space by exploiting the properties of a Euclidean model of
the space, such as the Beltrami-Klein or Poincar\'{e}. 

Hyperbolic and spherical space have also been used to display
self-organizing maps in the context of data visualization. Ontrup and
Ritter~\cite{ontrup-hyperbolic} and
Ritter~\cite{ritter99selforganizing} extend the traditional use of a
regular (Euclidean) grid, on which the self-organizing map is created,
with a tessellation in spherical or hyperbolic space.  An iterative
process is then used to adjust which elements in the data-set are
represented by the intersections.  Although the hyperbolic space
method seems a promising way to display high-dimensional data-sets,
the restriction to a lattice is often undesirable for graph
visualization.

Ostry~\cite{ostry96some} considers constraining force-directed
algorithms to the surface of three-dimensional objects. This work is
based on a differential equation formulation of the motion of the
nodes in the graph, and is flexible in that it allows a layout on
almost any object, even multiple objects.  Since the force
calculations are made in Euclidean space, however, this method is
inapplicable to certain geometries (e.g., hyperbolic geometry).

Another example of graph embedding within a non-Euclidean geometry is
described in the context of generating spherical parameterizations of
3D meshes. Gotsman {\em et al.}~\cite{ggs-fund-03} describe a method
for producing such an embedding using a generalization to spherical
space of planar methods for expressing convex combinations of points.
The implementation of the procedure is similar to the method described
in this paper, but it may not lend itself to geometries other than
spherical.

\begin{figure*}[t]
    \begin{center}
    \includegraphics[width=\textwidth]{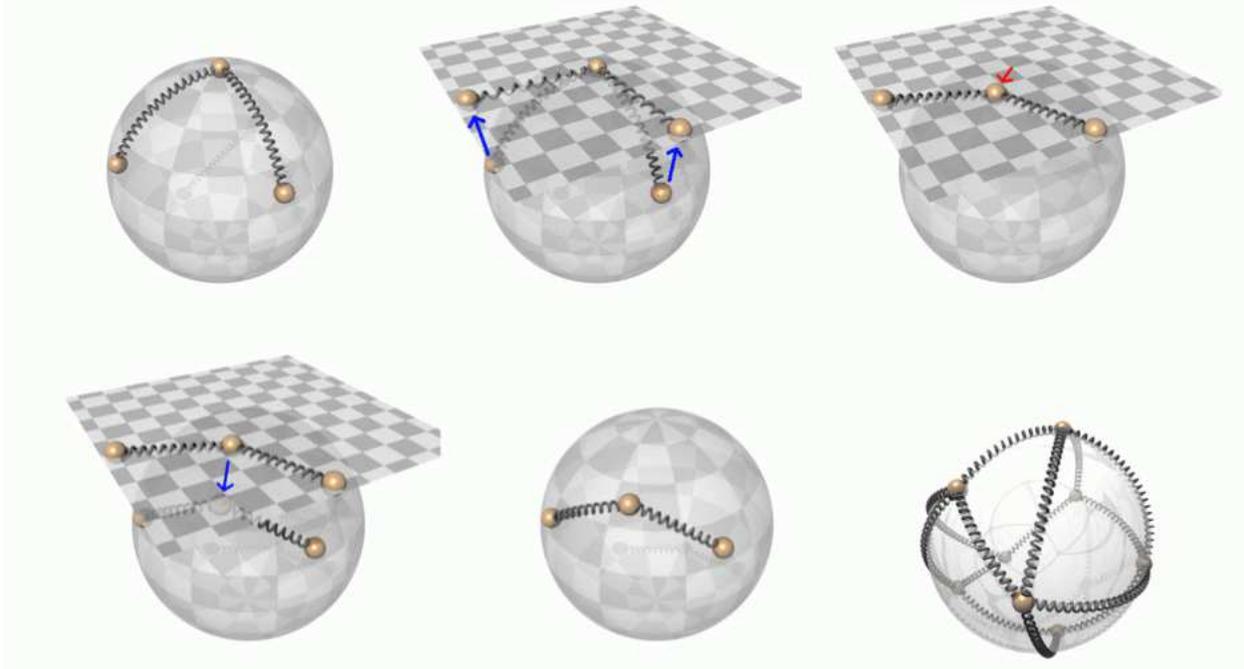} 
\end{center}
   \caption{\small\sf An overview of a spring embedder on the sphere.}
\label{fd:fig:Riem}
\end{figure*}

\begin{figure*}[t]
\includegraphics[width=.48\textwidth]{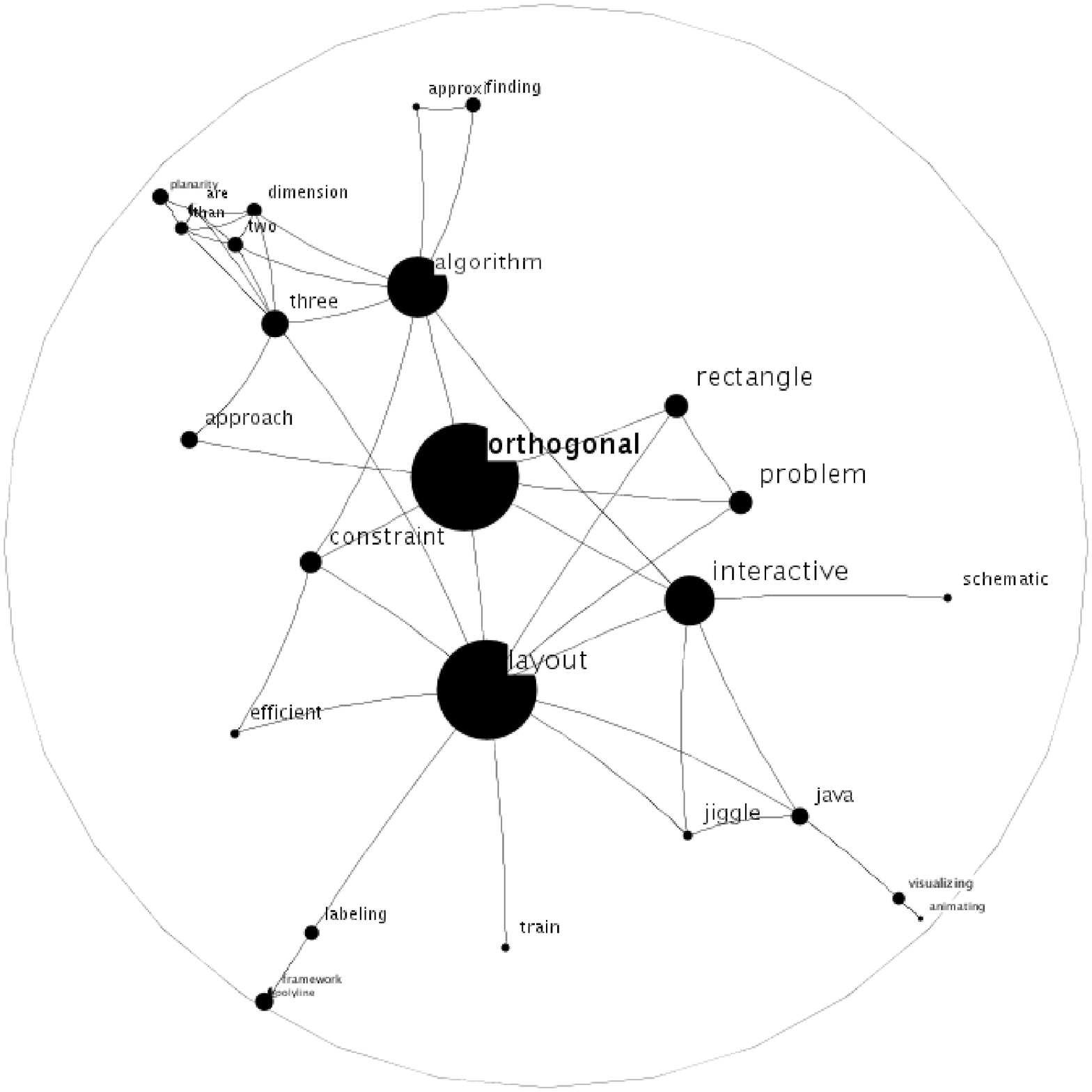}
    \includegraphics[width=.48\textwidth]{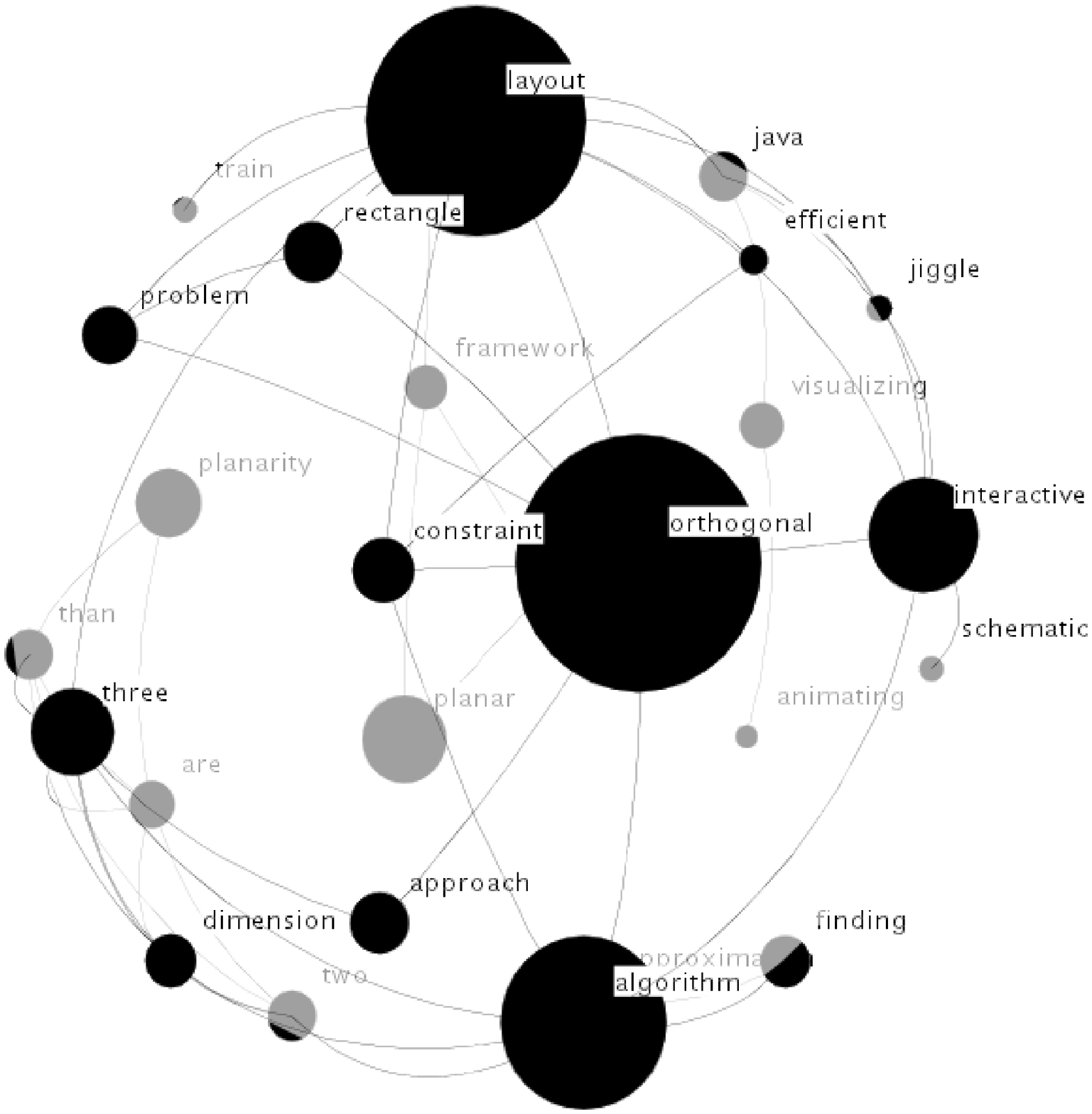}
\caption
    {\small\sf
Layouts of a graph obtained from research papers titles in hyperbolic
space $\mathbb{H}^2$ and in spherical space $\mathbb{S}^2$~\cite{kw-nese-05}.}
\label{fd:fig:sph}
\end{figure*}

Kobourov and Wampler~\cite{kw-nese-05} describe a conceptually simple
approach to generalizing force-directed methods for graph layout from
Euclidean geometry to Riemannian geometries; see Fig.~\ref{fd:fig:Riem}. Unlike previous work on
non-Euclidean force-directed methods, this approach is not limited to
special classes of graphs but can be applied to arbitrary graphs; see Fig~\ref{fd:fig:sph}. The
method relies on extending the Euclidean notions of distance, angle,
and force-interactions to smooth non-Euclidean geometries via
projections to and from appropriately chosen tangent spaces. Formal
description of the calculations needed to extend such algorithms to
hyperbolic and spherical geometries are also detailed.

In 1894 Riemann described a generalization of the geometry of
surfaces, which had been studied earlier by Gauss, Bolyai, and
Lobachevsky. Two well-known special cases of Riemannian geometries are
the two standard non-Euclidean types, spherical geometry and
hyperbolic geometry. This generalization led to the modern concept of
a Riemannian manifold. Riemannian geometries have less convenient
structure than Euclidean geometry, but they do retain many of the
characteristics which are useful for force-directed graph layouts.  A
Riemannian manifold $M$ has the property that for every point $x \in
M$, the tangent space $T_xM$ is an inner product space. This means
that for every point on the manifold, it is possible to define local
notions of length and angle.

Using the local notions of length we can define the length of a
continuous curve $\gamma : [a,b] \rightarrow M$ by

\begin{equation*}
    length(\gamma) = \int_a^b || \gamma'(t) || dt.
\end{equation*}

This leads to a natural generalization of the concept of a straight
line to that of a {\em geodesic}, where the geodesic between two
points $u,v\in M$ is defined as a continuously differentiable curve of
minimal length between them.  These geodesics in Euclidean geometry
are straight lines, and in spherical geometry they are arcs of great
circles.  We can similarly define the distance between two points,
$d(x,y)$ as the length of a geodesic between them. In Euclidean space
the relationship between a pair of nodes is defined along lines: the
distance between the two nodes is the length of the line segment
between them and forces between the two nodes act along the line
through them.  These notions of distance and forces can be extended to
a Riemannian geometry by having these same relationships be defined in
terms of the geodesics of the geometry, rather than in terms of
Euclidean lines.

As Riemannian manifolds have a well-structured tangent space at every
point, these tangent spaces can be used to generalize spring embedders
to arbitrary Riemannian geometries. In particular, the tangent space
is useful in dealing with the interaction between one point and
several other points in non-Euclidean geometries.  Consider three
points $x$, $y$, and $z$ in a Riemannian manifold $M$ where there is
an attractive force from $x$ to $y$ and $z$. As can be easily seen in
the Euclidean case (but also true in general) the net force on $x$ is
not necessarily in the direction of $y$ or $z$, and thus the natural
motion of $x$ is along neither the geodesic toward $y$, nor that
toward $z$. Determining the direction in which $x$ should move
requires the notion of angle.

Since the tangent space at $x$, being an inner product space, has
enough structure to define lengths and angles, we do the computations
for calculating the forces on $x$ in $T_x M$.  In order to do this, we
define two functions for every point $x \in M$ as follows:
\begin{eqnarray*}
 \fcn{\tau_x}{M}{T_x M} \\
 \fcn{\tau_x^{-1}}{T_x M}{M}.
 \end{eqnarray*}

These two functions map points in $M$ to and
from the tangent space of $M$ at $x$, respectively. We require that
$\tau_x$ and $\tau_x^{-1}$ satisfy the following constraints:

\begin{enumerate}
    \item $\tau_x^{-1}(\tau_x(y)) = y$ for all $y \in M$
    \item $|| \tau_x(y) || = d(x,y)$
    \item $\tau_x$ preserves angles about the origin
\end{enumerate}

Using these functions it is now easy to define the way in which the
nodes of a given graph $G=(V,E)$ interact with each other through
forces.  In the general framework for this algorithm each node is
considered individually, and its new position is calculated based on
the relative locations of the other nodes in the graph (repulsive
forces) and on its adjacent edges (attractive forces). Then we obtain
pseudo-code for a traditional Euclidean spring embedder and its
corresponding Riemannian counterpart; see Algorithm~\ref{alg:Riem}.

\begin{algorithm}
{\bf Euclidean}($G$) \tcc*[r]{Generic force-directed algorithm}
\For{$i=1$ \KwTo $Iterations$}{
\ForEach{$v \in V(G)$}{
$position[v] \gets force\_directed\_placement(v, G)$
}
}
 {\bf Riemannian}($G$) \tcc*[r]{non-Euclidean counterpart}
 \For{$i=1$ \KwTo $Iterations$}{
\ForEach{$v \in V(G)$}{
        $x \gets position[v]$\;
        $G' \gets \tau_x(G)$\;
        $x' \gets force\_directed\_placement(v, G')$\;
        $position[v] \gets \tau_x^{-1}(x')$\;
}
}
\caption{Euclidean and Riemannian\label{alg:Riem}}
\end{algorithm}

\section{Lombardi Spring Embedders}
\label{fd:sec:lom}

Inspired by American graphic artist Mark Lombardi, Duncan {\it et al.}~\cite{degkn-dtwpa-10,degkn-ldg-10}
introduce the concept of a \emph{Lombardi drawing}, which is a
drawing that uses circular arcs for edges and achieves the maximum
(i.e., \emph{perfect}) amount of angular resolution possible at each
vertex. 

There are several force-directed graph drawing methods
that use circular-arc edges or curvilinear poly-edges.
Brandes and Wagner~\cite{DBLP:journals/jgaa/BrandesW00}
describe a force-directed method for drawing train connections, where the vertex positions are fixed
but transitive edges are drawn as B{\'e}zier curves.
Finkel and Tamassia~\cite{ft-cgduf-05}, on the other hand,
describe a force-directed
method for drawing graphs using curvilinear edges where vertex
positions are free to move. Their method is
based on adding dummy vertices that serve as control points for B{\'e}zier curve.

\begin{figure*}[t]
\begin{center}
\includegraphics[width=.32\textwidth]{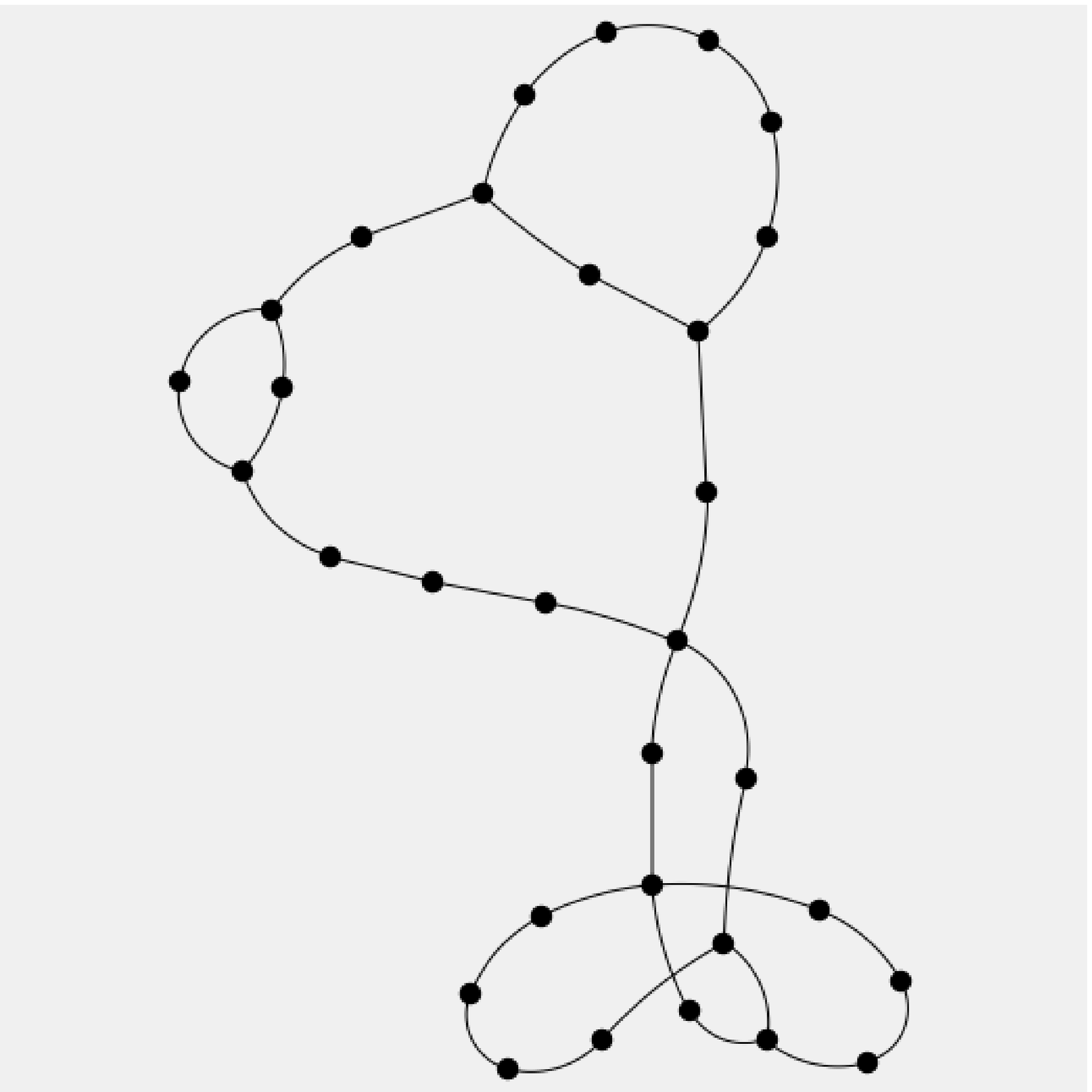}
\includegraphics[width=.32\textwidth]{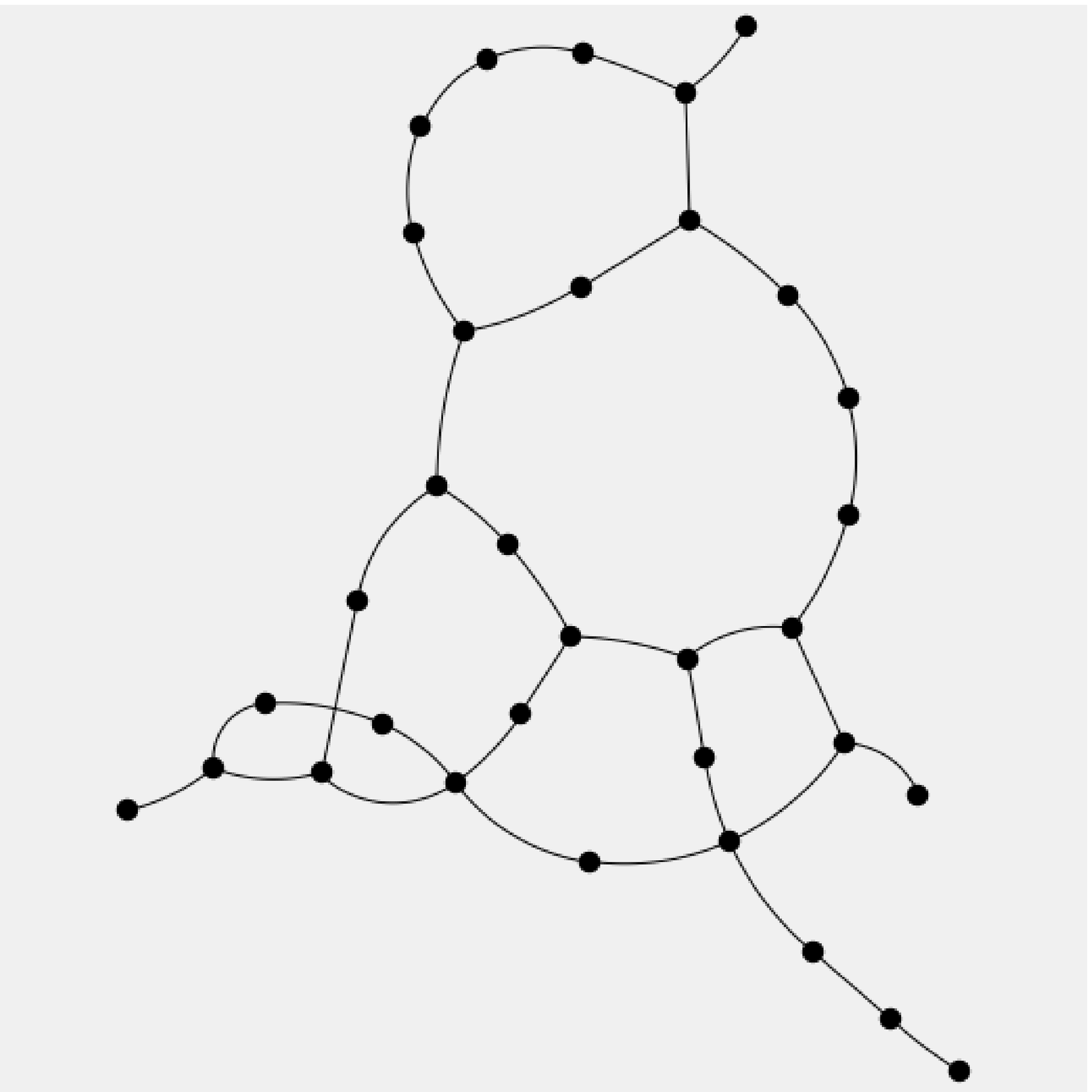}
\includegraphics[width=.32\textwidth]{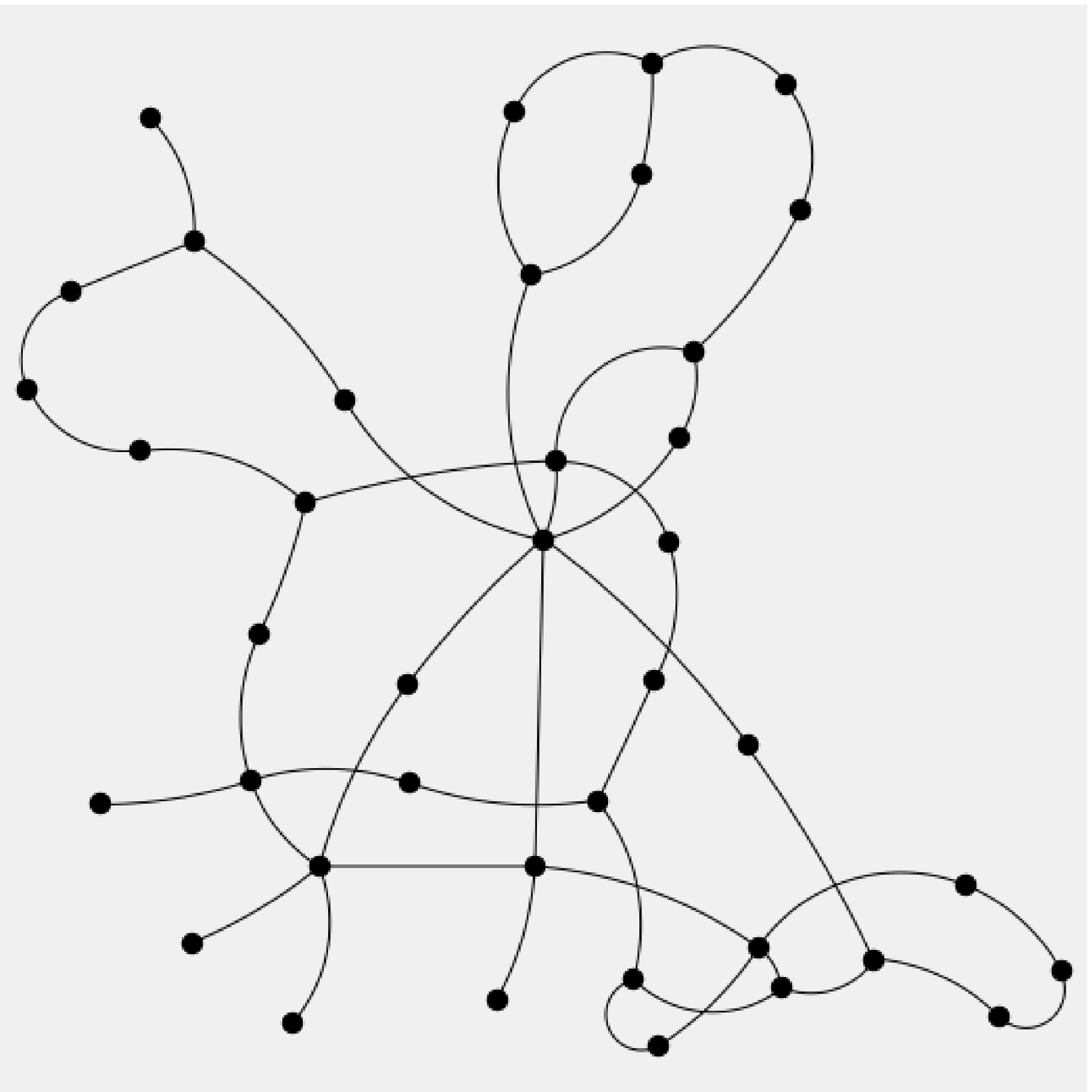}
\caption{\small\sf Examples of force-directed Lombardi drawings: note
  that every edge is a circular arc and every vertex has perfect
  angular resolution~\cite{ccgkt-fdl-11}.
\label{fd:fig:lom}
}
\end{center}
\end{figure*}

Chernobelskyi {\em et al.}~\cite{ccgkt-fdl-11} describe two force-directed algorithms for 
\emph{Lombardi-style} (or \emph{near-Lombardi})
drawings of graphs, where edges are drawn using circular arcs with
the goal of maximizing the angular resolution at each vertex.
The first approach calculates
lateral and rotational forces based on the two tangents
defining a circular arc between two vertices.
In contrast, the second approach uses
dummy vertices on each edge with repulsive forces
to ``push out'' the circular arcs representing
edges, so as to provide an aesthetic ``balance.''
Another distinction between the two approaches is that
the first one lays out the vertex positions along with
the circular edges, while the second one works on graphs
that are already laid out, only modifying the edges. It can be argued
that Lombardi or near-Lombardi graph drawings have a certain aesthetic
appeal; see Fig.~\ref{fd:fig:lom}. However, no experiments have been performed to test whether they
improve graph readability.

\section{Dynamic Graph Drawing}
\label{fd:sec:dyn}

While static graphs arise in many applications, dynamic processes give
rise to graphs that evolve through time. Such dynamic processes can be
found in software engineering, telecommunications traffic,
computational biology, and social networks, among others. 

\begin{figure*}[t]
\begin{center}
\includegraphics[width=.45\textwidth]{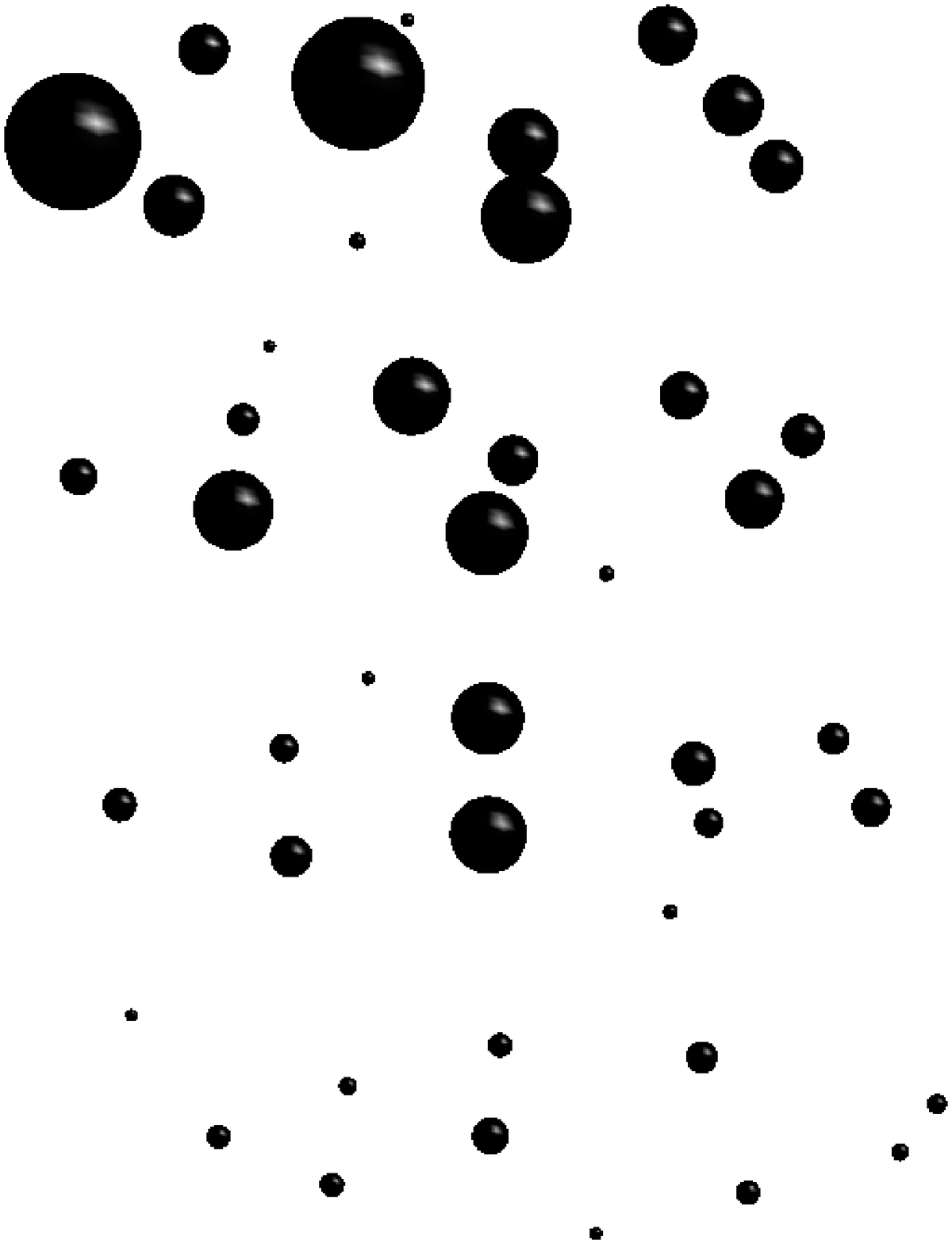}
\includegraphics[width=.45\textwidth]{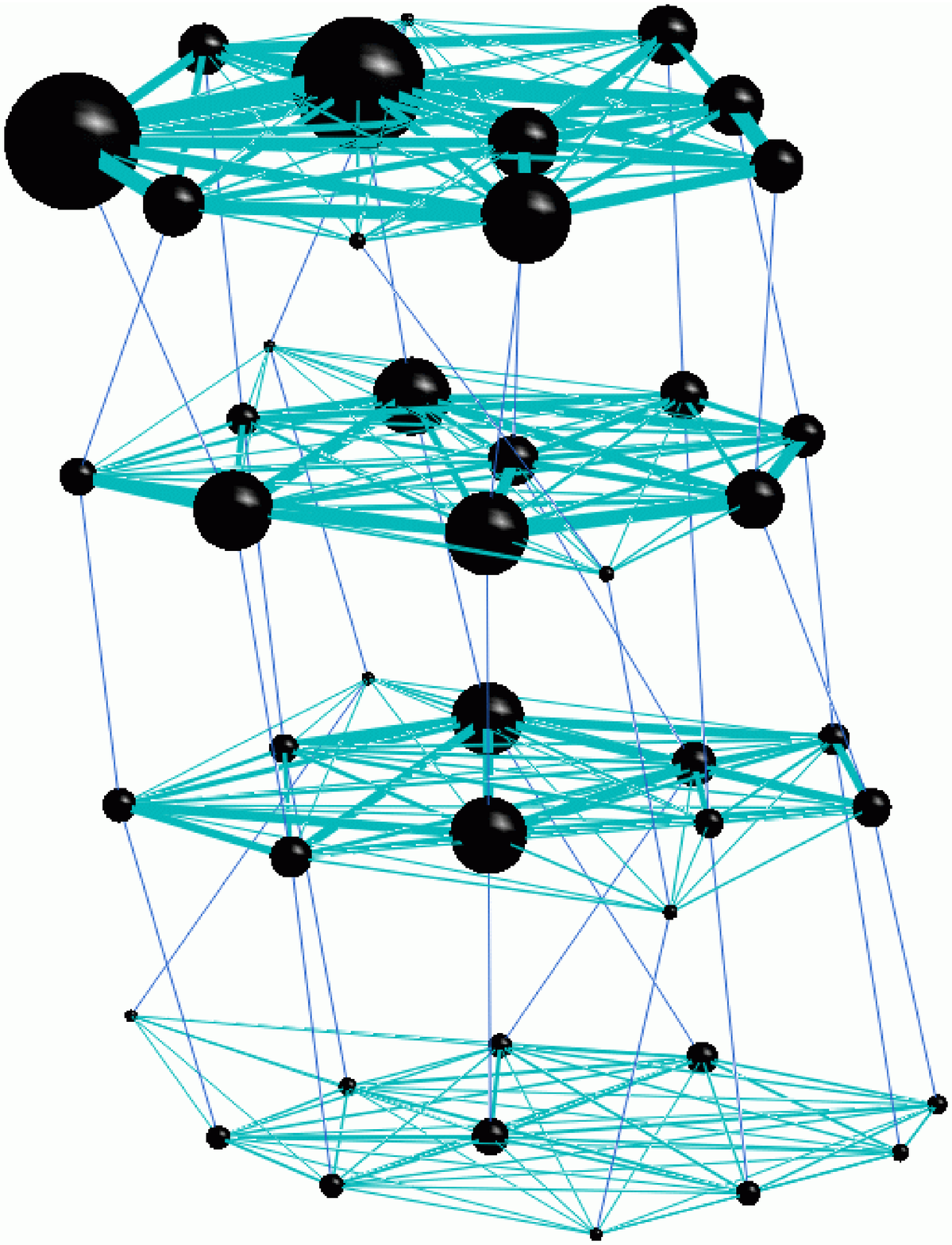}
\end{center}
\caption {\small\sf A dynamic graph can be interpreted as a larger graph made
  of connecting graphs in adjacent timeslices~\cite{ehkwy}.
}
\label{fd:fig:dyn}
\end{figure*}

Thus, dynamic graph drawing deals with the problem of effectively
presenting relationships as they change over time. A related problem is that of
visualizing multiple relationships on the same dataset. Traditionally,
dynamic relational data is visualized with the help of graphs, in
which vertices and edges fade in and out as needed, or as a time-series of
graphs; see Fig~\ref{fd:fig:dyn}.

The input to this problem is a series of graphs defined on the same underlying set of vertices. As a consequence, nearly all existing approaches to
visualization of evolving and dynamic graphs are based on the
force-directed method. Early work can be dated back to North's DynaDAG~\cite{North:1996:ILD},
where the graph is not given all at once, but incrementally. Brandes
and Wagner adapt the force-directed model to dynamic graphs using a
Bayesian framework [Brandes and Wagner 1998]. Diehl and
G\"org~\cite{Diehl:2002:GTC} consider graphs in a sequence to create
smoother transitions. Special classes of graphs such as trees,
series-parallel graphs and st-graphs have also been studied in dynamic
models~\cite{cdtt-dgdts-95,SCG92*261,Moen:1990:DDT}.	Most of these
early approaches, however, are limited to special classes of graphs
and usually do not scale to graphs over a few hundred vertices. 

\begin{figure}[t]
\begin{center} 
\includegraphics[width=\textwidth]{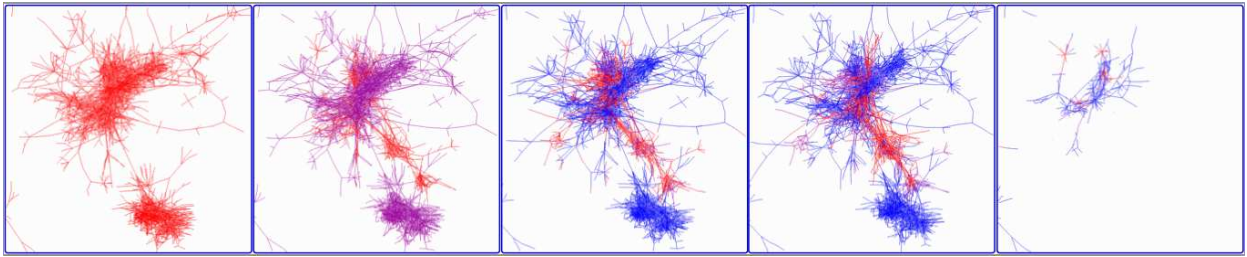}
\end{center}
\caption{\small\sf  
Snapshots of the call-graph of a program as it evolves through time, extracted from CVS logs. Vertices start out red. As time passes and a vertex does not change it turns purple and finally blue. When another change is affected, the vertex again becomes red. Note the number of changes between the two large clusters and the break in the build on the last image.~\cite{ck-sgbv-03}.}
\label{fig-callgraphs}
\end{figure}

{\tt TGRIP} was one of the first practical tools that could handle the larger graphs that
appear in the real-world. It was developed as part of a system that
keeps track of the evolution of software by extracting
information about the program stored within a CVS version control
system~\cite{ck-sgbv-03}. Such tools allow programmers to understand
the evolution of a legacy program: Why is the program structured the
way it is? Which programmers were responsible for which parts of the
program during which time periods? Which parts of the program appear
unstable over long periods of time? {\tt TGRIP} was used to visualize inheritance graphs, program call-graphs, and
control-flow graphs, as they evolve over time; see
Fig.~\ref{fig-callgraphs}.

For layout of evolving and dynamic graphs, there are two important
criteria to consider:
\begin{enumerate}
\item {\em readability} of the individual layouts, which 
depends on aesthetic criteria such
as display of symmetries, uniform edge lengths, and minimal number of
crossings; and
\item {\em mental map preservation} in the series of layouts, which can be achieved by ensuring
that vertices and edges that appear in consecutive graphs in the
series, remain in the same location. 
\end{enumerate}
These two criteria are often
contradictory. If we obtain individual layouts for each graph, without
regard to other graphs in the series, we may optimize readability at
the expense of mental map preservation. Conversely, if we fix the
common vertices and edges in all graphs once and for all, we are
optimizing the mental map preservation yet the individual layouts may
be far from readable. Thus, we can measure the effectiveness of
various approaches for visualization of evolving and dynamic graphs by
measuring the readability of the individual layouts, and the overall
mental map preservation.

Dynamic graphs can be visualized with {\em aggregated views}, where
all the graphs are displayed at once, {\em merged views}, where all the
graphs are stacked above each other, and with {\em animations}, where only
one graph is shown at a time, and morphing is used when changing
between graphs (fading in/out vertices and edges that
appear/disappear). When using the animation/morphing approach,
it is possible to change the balance between
readability of individual graphs and the overall mental map
preservation, as in the system for Graph Animations with Evolving
Layouts, GraphAEL~\cite{graphael03,fknwe-graphael-04}. Applications of
this framework include visualizing software evolution~\cite{ck-sgbv-03}, social
networks analysis~\cite{gephi}, and the behavior of dynamically modifiable
code~\cite{didfk-vbdmc-05}.

\section{Conclusion}
Force directed algorithms for drawing graphs have a long history and
new variants are still introduced every year. Their intuitive
simplicity appeals to researchers from many different fields, and this
accounts for dozens of available implementations. As new relational data
sets continue to be generated in many applications, force directed algorithms will likely
continue to be the method of choice. The latest scalable algorithms
and algorithms that can handle large dynamic and streaming graphs are
arguably of greatest utility today.

\bibliographystyle{abbrv}
\bibliography{stephen,fd}
\end{document}